\documentclass[rmp,aps,nofootinbib]{revtex4}
\usepackage{graphics}
\usepackage{epsfig}
\usepackage{amssymb}
\usepackage{float}
\usepackage{makeidx}
\usepackage{amsmath}

\def\inbar{\,\vrule height1.5ex width.4pt depth0pt}
\def\IR{\relax{\rm I\kern-.18em R}}
\def\IC{\relax\hbox{$\inbar\kern-.3em{\rm C}$}}

\begin{document}

\title{Electron transport through double quantum dots}

\author{W.~G. van der Wiel \vspace{-0.2cm}}
\email{wilfred@tomoko.phys.s.u-tokyo.ac.jp}
\affiliation{Department of Physics and ERATO Mesoscopic
Correlation Project, Tokyo University, 7-3-1 Hongo, Bunkyo-ku,
Tokyo 113-0033, Japan; Department of Applied Physics and DIMES,
Delft University of Technology, PO Box 5046, 2600 GA Delft, The
Netherlands}
\author{S. De Franceschi}
\affiliation{Department of Applied Physics, DIMES, and ERATO
Mesoscopic Correlation Project, Delft University of Technology, PO
Box 5046, 2600 GA Delft, The Netherlands}
\author{J.~M. Elzerman \vspace{-0.2cm}}
\affiliation{Department of Applied Physics, DIMES, and ERATO
Mesoscopic Correlation Project, Delft University of Technology, PO
Box 5046, 2600 GA Delft, The Netherlands}
\author{T. Fujisawa \vspace{-0.2cm}}
\affiliation{NTT Basic Research Laboratories, Atsugi-shi, Kanagawa
243-0198, Japan}
\author{S. Tarucha \vspace{-0.2cm}}
\affiliation{Department of Physics and ERATO Mesoscopic
Correlation Project, Tokyo University, 7-3-1 Hongo, Bunkyo-ku,
Tokyo 113-0033, Japan}
\author{L.~P. Kouwenhoven \vspace{-0.2cm}}
\affiliation{Department of Applied Physics, DIMES, and ERATO
Mesoscopic Correlation Project, Delft University of Technology, PO
Box 5046, 2600 GA Delft, The Netherlands~}
\vspace{-0.25cm}
\begin{abstract}
Electron transport experiments on two lateral quantum dots coupled
in series are reviewed. An introduction to the charge stability
diagram is given in terms of the electrochemical potentials of
both dots. Resonant tunneling experiments show that the double dot
geometry allows for an accurate determination of the intrinsic
lifetime of discrete energy states in quantum dots. The evolution
of discrete energy levels in magnetic field is studied. The
resolution allows to resolve avoided crossings in the spectrum of
a quantum dot. With microwave spectroscopy it is possible to probe
the transition from ionic bonding (for weak inter-dot tunnel
coupling) to covalent bonding (for strong inter-dot tunnel
coupling) in a double dot artificial molecule. This review on the
present experimental status of double quantum dot studies is
motivated by their relevance for realizing solid state quantum
bits.
\end{abstract}

\maketitle

\vspace{-1.5cm} \tableofcontents

\section{INTRODUCTION}
\label{sec:intro}

Quantum dots are man-made sub-micron structures in a solid,
typically consisting of 10$^3$-10$^9$ atoms and a comparable
number of electrons \cite{Kouwenhoven:1997}. In semiconductor
quantum dots all electrons are tightly bound, except for a small
number of free electrons, which can range from zero to several
thousands. For the quantum dot devices considered in this review,
the starting point for fabrication is formed by a heterostructure
consisting of different semiconductor materials (GaAs/AlGaAs). The
free electrons are strongly confined to the interface between GaAs
and AlGaAs, forming a 2-dimensional electron gas (2DEG).
Confinement in the other two dimensions is accomplished by locally
depleting the 2DEG, via etching techniques or metal gate
electrodes. The resulting structure is weakly coupled to source
and drain electrical contacts by tunnel
barriers, connecting it to the outside world.\\
\indent An important element of electronic transport through
quantum dots is Coulomb blockade
\cite{Kouwenhoven:1997,Grabert:1992,Averin:1986,Averin:1991}. An
extra electron can only be added to the dot, if enough energy is
provided to overcome the Coulomb repulsion between the electrons.
Next to this purely classical effect, the confinement in all three
directions leads to quantum effects that strongly influence
electronic transport at low temperature. In particular it leads to
the formation of a discrete (0D) energy spectrum, resembling that
of an atom. This and other similarities have therefore lead to the
name artificial
atoms for quantum dots \cite{Kastner:1993}.\\
\indent The next logical step after studying individual quantum
dots is to study systems of more than one dot. Where single
quantum dots are regarded as `artificial atoms', two quantum dots
can be coupled to form an `artificial molecule'. Depending on the
strength of the inter-dot coupling, the two dots can form
ionic-like (weak tunnel coupling) or covalent-like bonds (strong
tunnel coupling). In the case of ionic bonding the electrons are
localized on the individual dots. The binding occurs, because a
static redistribution of electrons leads to an attractive Coulomb
interaction. Weakly, electrostatically coupled quantum dots with
negligible inter-dot tunnel conductance are covered by orthodox
Coulomb blockade theory \cite{Averin:1991}. In the case of
covalent bonding, two electron states are quantum-mechanically
coupled. The main requirement for covalent binding is that an
electron can tunnel many times between the two dots in a
phase-coherent way. Here the electron cannot be regarded as a
particle that resides in one particular dot, but it must be
thought of as a coherent wave that is delocalized over the two
dots. The bonding state of a strongly coupled artificial molecule
has a lower energy than the energies of the original states of the
individual dots. This
energy gain forms the binding force between the two dots.\\
\indent The theoretical possibility to perform certain tasks in a
much more efficient way using a `quantum computer' instead of a
`classical computer', has stimulated the search for physical
realizations of the basic building block of such a computer: the
quantum bit. In principle, any quantum two-level system can be
used as such a qubit. In particular, recent studies have put
forward double quantum dots as interesting candidates for
realizing qubits \cite{Loss:1998}. The possible application of
double quantum dot devices in quantum logic forms an important
motivation
for this work.\\
\indent In this review we concentrate on electron transport
through lateral double quantum dots coupled in series. All devices
have been fabricated and all experiments have been performed at
Delft University of Technology and NTT Basic Research
Laboratories. By now there exists an extensive literature on
experimental studies of electron transport through double lateral
quantum dots coupled in series\footnote{Electron transport
measurements in double lateral quantum dots coupled in series are
described in
\cite{Blick:1996,Blick:1998,Dixon:1996,Dixon:1998,Fujisawa:1996,Fujisawa:1997a,Fujisawa:1997b,Fujisawa:1998,Ishibashi:1998,Jeong:2001,Kemerink:1994,Livermore:1996,Molenkamp:1995,Oosterkamp:1998a,Oosterkamp:1998b,Vaart:1995,Waugh:1995,Waugh:1996}},
and lateral double dots coupled in parallel\footnote{Electron
transport measurements in double lateral quantum dots coupled in
parallel are described in
\cite{Adourian:1996,Adourian:1999,Dixon:1998,Hofmann:1995}}.
Vertical double quantum dot structures\footnote{Electron transport
measurements in vertical double quantum dots are described in
\cite{Reed:1989,Tewordt:1992,Schmidt:1997,Austing:1998,Tarucha:1999,Austing:2001,Amaha:2001}}
fall outside the focus of this review. In vertical structures, the
characteristics of the tunnel barriers are set by the growth
parameters of the heterostructure, limiting the experimental
tunability. Besides that, the gate geometry used in these devices,
makes it difficult to address dots independently.\\
\indent As a first step to understanding double dot systems we
introduce the stability diagram \cite{Pothier:1992}, or honeycomb
diagram, in section \ref{stability}. It is a convenient tool in
the analysis of double dot transport properties. Resonant
tunneling experiments discussed in section \ref{restun}, show that
the resonant widths are only determined by the lifetime of the
discrete energy states, independent of the electron temperature.
In section \ref{magnetiz} we discuss level spectroscopy in a
magnetic field. The double dot geometry offers sufficient energy
resolution to probe intra-dot level repulsion. In section
\ref{spectro} we present microwave spectroscopy measurements on a
quantum dot molecule. We illustrate the transition from a weakly
coupled double dot to a strongly coupled double dot, by discussing
a two-level system in section \ref{2level}. Although being a clear
simplification, the mapping of the double dot on a two-level
system grasps much of the physics of the experiments presented in
section \ref{spectro}. Irradiation with microwaves leads to photon
assisted tunneling (PAT) (sections \ref{weakcPAT} and
\ref{strongcPAT}), which turns out to be a powerful tool not only
to reveal the character of the inter-dot coupling, but also to
quantitatively determine the bonding strength.

\section{Stability diagram}
\label{stability}

In this section we introduce the stability -- or honeycomb --
diagram that visualizes the equilibrium charge states of two
serially coupled dots.

\subsection{Linear transport regime}

\label{linear}

\subsubsection{Classical theory}

\label{linclass}

\begin{figure}[htbp]
  \begin{center}
    \centerline{\epsfig{file=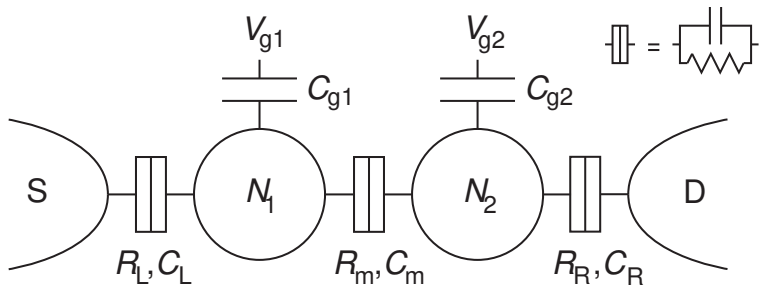, width=10cm, clip=true}}
    \caption{Network of resistors and capacitors representing two quantum dots coupled in series.
    The different elements are explained in the text. Note that tunnel barriers are characterized
    by a resistor and a capacitor, as indicated in the inset.}
    \label{network}
  \end{center}
\end{figure}

We start with a purely classical description in which the
influence of discrete quantum states is not taken into account yet
\cite{Dixon:1998,Pothier:1992,Ruzin:1992}\footnote{Ruzin {\it et
al.} have an eror in their Eq$.$ (9), namely the numerator, ($2C +
C_i \delta_{ij}$). For $i=j$, this relation gives ($2C + C_1$)
when it should read ($2C + C_2$), and vice-versa.}. The double dot
is modeled as a network of resistors and capacitors (Fig$.$
\ref{network}). The number of electrons on dot 1(2) is $N_{1(2)}$.
Each dot is
capacitively coupled to a gate voltage $V_{g1(2)}$ through a
capacitor $C_{g1(2)}$ and to the source (S) or drain (D) contact
through a tunnel barrier represented by a resistor $R_{L(R)}$ and
a capacitor $C_{L(R)}$ connected in parallel. The dots are coupled
to each other by a tunnel barrier represented by a resistor $R_m$
and a capacitor $C_m$ in parallel. The bias voltage, $V$, is
applied to the source contact with the drain contact grounded
(asymmetric bias). In this section we consider the linear
transport regime, i$.$e$.$ $V \approx 0$. If cross-capacitances
(such as between $V_{g1}$ and dot 2), other voltage sources and
stray capacitances are negligible, the double dot electrostatic
energy reads (a full derivation is given in the appendix)

\begin{align}
U(N_1,N_2) = & \hspace{0.2cm} \frac{1}{2} N_1^2 E_{C1} +
\frac{1}{2} N_2^2 E_{C2}
+ N_1 N_2 E_{Cm} + f(V_{g1},V_{g2}) \label{eqDDenergy} \\
f(V_{g1},V_{g2}) = & \hspace{0.2cm} \frac{1}{-|e|}\{C_{g1}
V_{g1}(N_1 E_{C1} + N_2 E_{Cm}) +  C_{g2} V_{g2}(N_1 E_{Cm} + N_2
E_{C2}) \} \nonumber
\\
& \hspace{0.2cm} + \frac{1}{e^2}\{\frac{1}{2} C^2_{g1}
V^2_{g1}E_{C1} + \frac{1}{2} C^2_{g2} V^2_{g2} E_{C2} + C_{g1}
V_{g1} C_{g2} V_{g2} E_{Cm} \} \nonumber
\end{align}
\\

\noindent where $E_{C1(2)}$ is the charging energy of the
individual dot 1(2) and $E_{Cm}$ is the electrostatic coupling
energy. The coupling energy $E_{Cm}$ is the change in the energy
of one dot when an electron is added to the other dot. These
energies can be expressed in terms of the capacitances as follows

\begin{equation}
E_{C1}=\frac{e^2}{C_1} \biggl( \frac{1}{1 - \frac{C_m^2}{C_1 C_2}}
\biggr); \ E_{C2}=\frac{e^2}{C_2} \biggl( \frac{1}{1 -
\frac{C_m^2}{C_1 C_2}} \biggr); \ E_{Cm}=\frac{e^2}{C_m} \biggl(
\frac{1}{\frac{C_1 C_2}{C_m^2}-1} \biggr) \label{eq2}
\end{equation}
\\

\noindent Here $C_{1(2)}$ is the sum of all capacitances attached
to dot 1(2) including $C_m$: $C_{1(2)} = C_{L(R)} + C_{g1(2)} +
C_m$. Note that $E_{C1(2)}$ can be interpreted as the charging
energy of the single, uncoupled dot 1(2) multiplied by a
correction factor that accounts for the coupling. When $C_m$ = 0,
and hence $E_{Cm}=0$, Eq$.$ (\ref{eqDDenergy}) reduces to

\begin{equation}
U(N_{1},N_{2})=\frac{(-N_1 |e| + C_{g1}V_{g1})^2}{2 C_1} +
\frac{(-N_2 |e| + C_{g2}V_{g2})^2}{2 C_{2}} \label{eq3}
\end{equation}
\\

\noindent This is the sum of the energies of two independent dots.
In the case when $C_m$ becomes the dominant capacitance
($C_m/C_{1(2)} \rightarrow 1$), the electrostatic energy is given
by

\begin{equation}
U(N_{1},N_{2})=\frac{\left[-(N_1 + N_2)|e| +
C_{g1}V_{g1}+C_{g2}V_{g2}\right]^2}{2(\widetilde{C}_{1}+\widetilde{C}_{2})}
\label{eq4}
\end{equation}
\\

\noindent This is the energy of a single dot with a charge $N_1 +
N_2$ and a capacitance of $\widetilde{C}_{1}+\widetilde{C}_{2}$,
where $\widetilde{C}_{1(2)}=C_{1(2)}-C_{m}$ is the capacitance of
dot 1(2) to the outside world. Thus, a large inter-dot capacitance
$C_m$ effectively leads to one big dot.\\
\indent The electrochemical potential $\mu _{1(2)}(N_{1},N_{2})$
of dot 1(2) is defined as the energy needed to add the
$N_{1(2)}$th electron to dot 1(2), while having $N_{2(1)}$
electrons on dot 2(1). Using the expression for the total energy
Eq$.$ (\ref{eqDDenergy}), the electrochemical potentials of the
two dots are

\begin{align}
\mu _{1}(N_{1},N_{2}) & \equiv U(N_{1},N_{2})-U(N_{1}-1,N_{2}) \nonumber \\
& =(N_{1}-\frac{1}{2})E_{C1}+N_{2}E_{Cm} - \frac{1}{|e|}\left(
C_{g1}V_{g1}E_{C1}+C_{g2}V_{g2}E_{Cm}\right) \label{eqmu1}
\end{align}

\begin{align}
\mu _{2}(N_{1},N_{2}) & \equiv U(N_{1},N_{2})-U(N_{1},N_{2}-1) \nonumber \\
& =(N_{2}-\frac{1}{2})E_{C2}+N_{1}E_{Cm} - \frac{1}{|e|}\left(
C_{g1}V_{g1}E_{Cm}+C_{g2}V_{g2}E_{C2}\right) \label{eqmu2}
\end{align}
\\

\noindent The change in $\mu_1(N_1,N_2)$ if, at fixed gate
voltages, $N_{1}$ is changed by one, $\mu _1 (N_1 +1, N_2) - \mu
_{1}(N_{1},N_{2}) = E_{C1}$, is called the {\it addition energy}
of dot 1 and equals the charging energy of dot 1 in this classical
regime. Similarly, the addition energy of dot 2 equals $E_{C2}$,
and $\mu _1 (N_1 , N_2 + 1) - \mu _{1}(N_{1},N_{2}) = \mu _2 (N_1
+ 1 , N_2 ) - \mu _{2}(N_{1},N_{2}) = E_{Cm}$. In the next section
we will discuss the addition energy in the quantum regime, where
also the spacing between discrete energy levels plays a role.\\
\indent From the electrochemical potentials in Eqs$.$
(\ref{eqmu1}) and (\ref{eqmu2}) we construct a charge stability
diagram, giving the equilibrium electron numbers $N_1$ and $N_2$
as a function of $V_{g1}$ and $V_{g2}$. We define the
electrochemical potentials of the left and right leads to be zero
if no bias voltage is applied, $\mu_L$ = $\mu_R$  = 0. Hence, the
equilibrium charges on the dots are the largest values of $N_1$
and $N_2$ for which both $\mu_1(N_1,N_2)$

\begin{figure}[htbp]
  \begin{center}
    \centerline{\epsfig{file=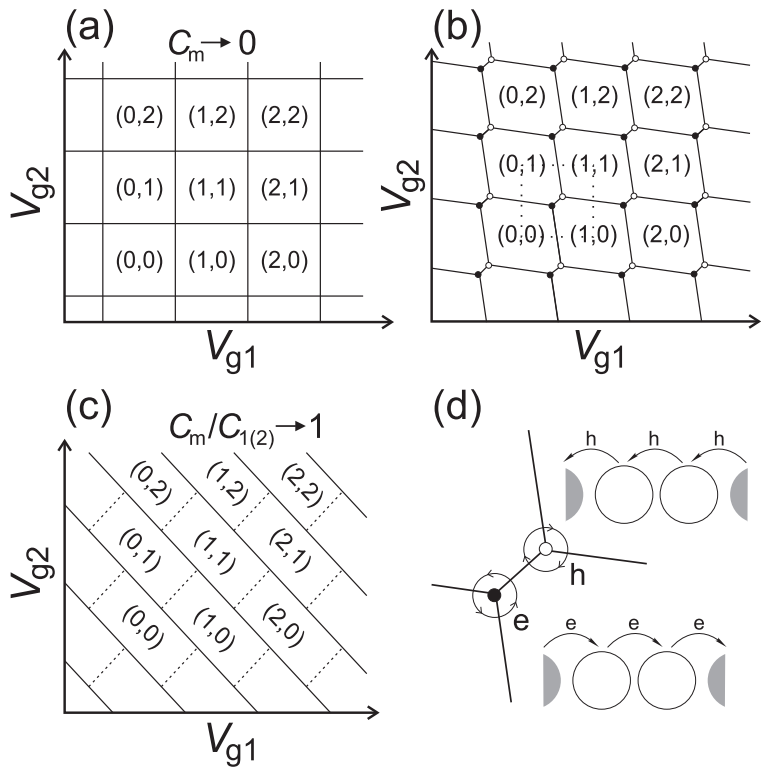, width=10cm, clip=true}}
    \caption{Schematic stability diagram of the double-dot system for (a) small, (b) intermediate, and (c) large
    inter-dot coupling. The equilibrium charge on each dot in each domain is denoted by ($N_1,N_2$).
    The two kinds of triple-points corresponding with the electron transfer process ({\Large $\bullet$})
    and the hole transfer process ({\Large $\circ$}) are illustrated in (d). The region in the dotted square in (b) is
    depicted in more detail in Fig$.$ \ref{honeyelch}.}
    \label{honeycombs}
  \end{center}
\end{figure}

\noindent and $\mu_2(N_1,N_2)$ are less than zero. If either is
larger than zero, electrons escape to the leads. This constraint,
plus the fact that $N_1$ and $N_2$ must be integers, creates
hexagonal domains in the $(V_{g1},V_{g2})$-phase space in which
the charge configuration is stable.\\
\indent For completely decoupled dots ($C_m$ = 0) the diagram
looks as in Fig$.$ \ref{honeycombs}(a). The gate voltage
$V_{g1(2)}$ changes the charge on dot 1(2), without affecting the
charge on the other. If the coupling is increased, the domains
become hexagonal (Fig$.$ \ref{honeycombs}(b)). The vertices of the
square domains have separated into `triple-points'. When $C_m$
becomes the dominant capacitance ($C_m/C_{1(2)} \rightarrow 1$),
the triple-point separation reaches its maximum (see Fig$.$
\ref{honeycombs}(c)). The double dot behaves like one dot with
charge $N_1 + N_2$, as seen from Eq$.$ (\ref{eq4}).
\\
\indent We are considering the linear regime of conductance,
implying $\mu_L - \mu_R = -|e|V \approx 0$. In order to obtain a
measurable current, the tunnel barriers need to be sufficiently
transparent. At the same time, however, the tunnel barriers need
to be sufficiently opaque to ensure a well-defined electron number
on each dot. For double dots coupled in series, a conductance
resonance is found when electrons can tunnel through both dots.
This condition is met whenever three charge states become
degenerate, i$.$e$.$ whenever three boundaries in the honeycomb
diagram meet in one point. In Fig$.$ \ref{honeycombs}(d) two kinds
of such triple-points are distinguished, ({\Large $\bullet$}) and
({\Large $\circ$}), corresponding to different charge transfer
processes. At the triple point ({\Large $\bullet$}), the dots
cycle through the sequence $(N_1,N_2) \rightarrow (N_1+1,N_2)
\rightarrow (N_1,N_2+1) \rightarrow (N_1,N_2)$, which shuttles one
electron through the system. This process is illustrated by the
counterclockwise path $e$ and the diagram of an electron
sequentially tunneling from the left lead to the right in Fig$.$
\ref{honeycombs}(d). At the other triple-point ({\Large $\circ$}),
the sequence is $(N_1+1,N_2+1) \rightarrow (N_1+1,N_2) \rightarrow
(N_1,N_2+1) \rightarrow (N_1+1,N_2+1)$, corresponding to the
clockwise path $h$ in Fig$.$ \ref{honeycombs}(d). This can be
interpreted as the sequential tunneling of a hole in the direction
opposite to the electron. The energy difference between both
processes determines the separation between the triple-points
({\Large $\bullet$}) and ({\Large $\circ$}), and is given by
$E_{Cm}$, as defined in Eq$.$ (\ref{eq2}).
\begin{figure}[htbp]
  \begin{center}
    \centerline{\epsfig{file=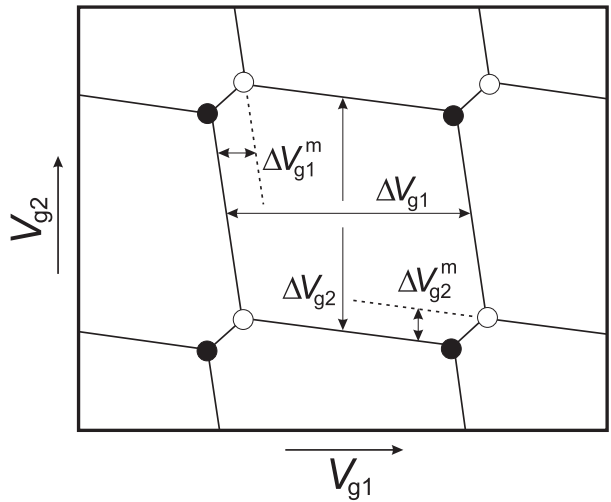, width=8cm, clip=true}}
    \caption{Schematic stability diagram showing the Coulomb peak
    spacings given in Eqs$.$ (\ref{eqDeltaVg}) and (\ref{eqDeltaVgm}).
    These spacings can be determined experimentally by connecting
    the triple-points.}
    \label{spacings}
  \end{center}
\end{figure}

The dimensions of the honeycomb cell (see Fig$.$ \ref{spacings})
can be related to the capacitances using Eqs$.$ (\ref{eqmu1}) and
(\ref{eqmu2}). From
\begin{equation}
\mu _1 (N_1,N_2;V_{g1},V_{g2}) = \mu _1 (N_1 +1,N_2;V_{g1} +
\Delta  V_{g1},V_{g2})
\end{equation}
we obtain
\begin{equation}
\Delta V_{g1}=\frac{|e|}{C_{g1}} \label{eqDeltaVg}
\end{equation}
and similarly we can derive
\begin{equation}
\Delta V_{g2}=\frac{|e|}{C_{g2}} \tag{\ref{eqDeltaVg}$'$}
\end{equation}
From
\begin{equation}
\mu _1 (N_1,N_2;V_{g1},V_{g2}) = \mu _1 (N_1,N_2+1;V_{g1} + \Delta
V_{g1}^m,V_{g2})
\end{equation}
we obtain
\begin{equation}
\Delta V_{g1}^{m}=\frac{|e| C_{m}}{C_{g1}C_{2}} = \Delta V_{g1}
\frac{C_m}{C_2} \label{eqDeltaVgm}
\end{equation}
and similarly we can derive
\begin{equation}
\Delta V_{g2}^{m}=\frac{|e| C_{m}}{C_{g2}C_{1}} = \Delta V_{g2}
\frac{C_m}{C_1} \tag{\ref{eqDeltaVgm}$'$}
\end{equation}
However, for a full characterization of all capacitances in the
system an analysis in the non-linear transport regime is required,
as is discussed in section \ref{nonlinear}.\\
\indent Above we assumed that $V_{g1}$ and $V_{g2}$ only couple
directly to the respective dots. In practice, however, there is a
finite cross-capacitance from one gate to the other. The
respective cross-capacitances result in a change of the slope of
the charge domain boundaries in the honeycomb diagram. From
Figs$.$ \ref{honeycombs}(b) and \ref{spacings} it is clear that
both kinds of triple points ({\Large $\bullet$} and {\Large
$\circ$}) form a square lattice. However, with finite
cross-capacitances the positions of the triple points move to
lower $V_{g1(2)}$ for increasing $V_{g2(1)}$ (at constant $V$).

\subsubsection{Quantized states}

\label{linquant}

The discussion of the stability diagram so far has been completely
classical. However, the strong confinement of electrons in the
dots can lead to the formation of a discrete energy spectrum. To
account for the quantized energy states in the dot, we need to
incorporate their energies in the electrochemical potential. The
electrochemical potential for adding an electron into energy level
$n$ of dot $i$ is denoted by $\mu_{i,n}$. Within the constant
interaction model, $\mu_{i,n}$ is the sum of the classical
electrochemical potential $\mu_i^{class}$ and the single-particle
energy $E_n$: $\mu_{i,n} = \mu_i^{class} + E_n$. In the classical
regime we found that the addition energy (the change in
electrochemical potential needed to add an extra electron) equals
the charging energy $E_{C1}$ (for dot 1) or $E_{C2}$ (for dot 2).
In the quantum regime, the addition energy for the $(N_{1}+1)$th
electron occupying discrete level $m$, with the $N_{1}$th electron
occupying discrete level $n$, becomes
\\
\begin{align}
\mu_{1,m}(N_1+1,N_2) - \mu_{1,n}(N_1,N_2) = & \hspace{0.1cm}
E_{C1} + (E_m - E_n) \nonumber \\
= & \hspace{0.1cm} E_{C1} + \Delta E.
\end{align}
\\
\noindent Similarly, we find $E_{C2} + \Delta E$ for the addition
energy of dot 2. Note that for a (spin-)degenerate level $\Delta
E$ can be zero. The dimensions of the honeycomb cell as given in
Eqs$.$ (\ref{eqDeltaVg}) and (\ref{eqDeltaVgm}), and depicted in
Fig$.$ \ref{spacings} for the classical regime, change as follows

\begin{align}
\Delta V_{g1(2)} = & \frac{|e|}{C_{g1(2)}} \biggl(1 + \frac{\Delta
E}{E_{C1(2)}} \biggr) \\
\nonumber \\
\Delta V_{g1(2)}^m = & \frac{|e|C_m} { C_{g1(2)} C_{2(1)} }
\biggl(1 + \frac{\Delta E}{E_{Cm}} \biggr)
\end{align}
\\

The electronic configuration that gives the lowest possible total
energy in dot 1(2), is referred to as the dot 1(2) {\it ground
state}. Any configuration with a higher total energy is referred
to as an {\it excited} state. The electrochemical potential for
adding the $N_{1(2)}$th electron to the lowest unfilled energy
level of the $(N_{1(2)}-1)$-electron ground state is labeled
$\mu_{1,0}(N_1,N_2)$ \{$\mu_{2,0}(N_1,N_2)$\}. The electrochemical
potential for adding the $N_{1(2)}$th electron to a higher
unfilled level of the $(N_{1(2)}-1)$-electron ground state -- or
to any unfilled level of an $(N_{1(2)}-1)$-electron excited state
-- is labeled $\mu_{1,1}(N_1,N_2)$, $\mu_{1,2}(N_1,N_2)$, ...
\{$\mu_{2,1}(N_1,N_2)$, $\mu_{2,2}(N_1,N_2)$, ...\}.\\
\begin{figure}[htbp]
  \begin{center}
    \centerline{\epsfig{file=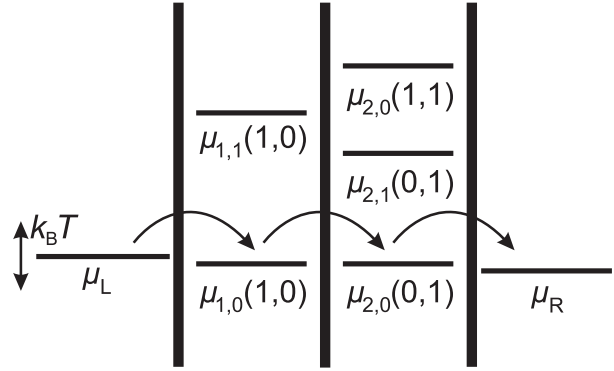, width=8cm, clip=true}}
    \caption{Schematic diagram of the electrochemical potentials $\mu_{i,n}(N_1,N_2)$ in dots and
    leads in the linear regime. The first subscript indicates either the lead (L,R) or the dot (1,2).
    The second subscript refers to the nature of the dot energy state (ground state, $n=0$ or $n$th
    excited state.}
    \label{linear_regime}
  \end{center}
\end{figure}

In Fig$.$ \ref{linear_regime} a schematic diagram is given,
showing the electrochemical potentials in the leads and dots in
the linear regime ($\mu_L - \mu_R = -|e|V \approx 0$). The
\begin{figure}[htbp]
  \begin{center}
    \centerline{\epsfig{file=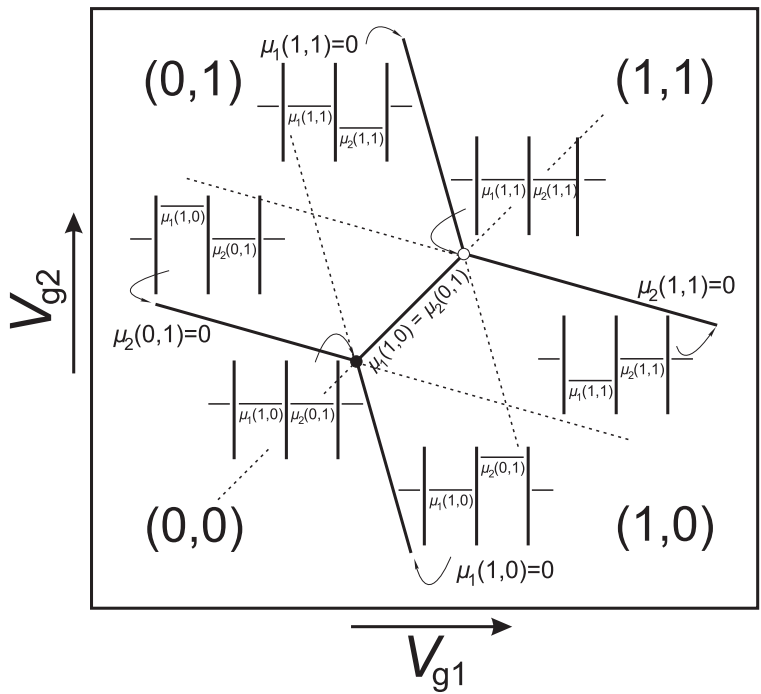, width=10cm, clip=true}}
    \caption{Region within the dotted square of Fig$.$ \ref{honeycombs}(b), corresponding with the `unit cell' of
    the double-dot stability diagram. Four different charge states can be distinguished, separated by solid
    lines. At the solid line connecting the two triple points, the charge
    states (0,1) and (1,0) are degenerate. At the other solid lines the electrochemical potential of at least
    one dot is zero and thus equals the electrochemical potential of the leads. The dashed lines are the
    extensions of the solid lines within the honeycomb cells. The triple-points lie on the
    crossing points between the solid lines. The schematic diagrams show the configuration of
    the ground-state electrochemical potentials on the corresponding
    place in the honeycomb diagram.}
    \label{honeyelch}
 \end{center}
\end{figure}
ground state electrochemical potentials $\mu_{1,0}(1,0)$ and
$\mu_{2,0}(0,1)$ align within the small bias window, allowing an
electron to tunnel from left to right. This is an example of an
electron transfer process as depicted in Fig$.$
\ref{honeycombs}(d). Note that an alignment of an arbitrary
combination of electrochemical potentials in dot 1 and dot 2 does
not necessarily lead to a current. For example, the alignment of
$\mu_{1,0}(1,0)$ and $\mu_{2,0}(1,1)$ does not result in current
through the double dot. In the linear regime electron transport
occurs via ground states, whereas the excited states start to play
a role in non-linear transport, as will be discussed in section
\ref{nonlinear}. In the following discussion of the linear regime
the ground state of dot 1(2) is denoted by $\mu_{1(2)}(N_1,N_2)$
(without the discrete level index).\\
\indent A more detailed picture of a honeycomb cell in the linear
regime, marked by the dashed square in Fig$.$ \ref{honeycombs}(b),
is given in Fig$.$ \ref{honeyelch}. The configuration of the
ground-state electrochemical potentials is given in schematic
diagrams on some places in the stability diagram. The dashed
lines, which are extensions of the solid lines forming the
honeycomb cells, help to find the position of both electrochemical
potentials on a certain place. A crossing of dashed lines (as in
the charge domains (0,1) and (1,0)) indicates that two
electrochemical potentials align, but does not result in a current
through the double dot.

\subsubsection{Experimental stability diagrams}

\label{stabexpchar}

Before discussing some experimental stability diagrams, we
introduce the two kinds of lateral double dot devices being
studied in this review. For the first type, only metal gate
electrodes are used to confine the electrons in the 2DEG beneath.
For the second type we use a combination of metal gates and dry
etching to realize confinement.

\begin{figure}[htbp]
  \begin{center}
    \centerline{\epsfig{file=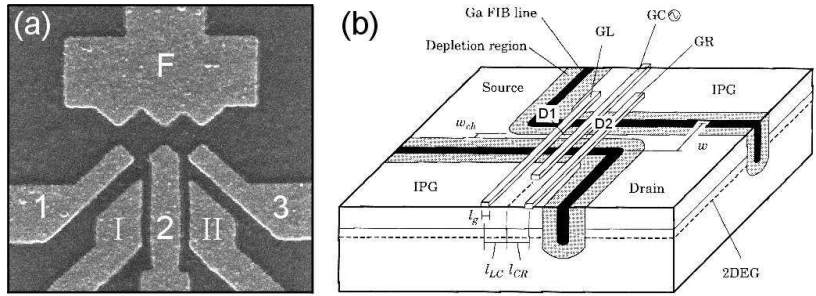, width=10.75cm, clip=true}}
    \caption{Double quantum dot devices. (a) SEM micrograph of a double dot defined by metallic
    gates (light gray areas). The ungated 2DEG (100 nm below the surface) has a mobility of 2.3 $\times$ $10^6$
    cm$^2$/(Vs)
    and an electron density of 1.9 $\times$ 10$^{15}$ m$^{-2}$ at 4.2 K. The dimensions
    of the dots defined by the gate pattern are 320 $\times$ 320 nm$^2$ (dot 1, left) and 280 $\times$ 280 nm$^2$ (dot 2, right)
    (b) Schematic diagram of a double dot defined
    by a combination of dry etching and metallic gates. The carrier concentration and mobility of the
    ungated 2DEG (100 nm below the surface) at 1.6 K and in the dark are 3 $\times$ 10$^{11}$ cm$^{-2}$
    and 8 $\times$ 10$^5$ cm$^2$/(Vs), respectively. The lithographic distance between the ethched
    trenches (black lines), $w$, is typically 0.5 $\mu$m. The effective width of the channel, $w_{ch}$, can be tuned by
    voltages on the in-plane gates (IPGs). The gate electrodes are $\sim$40 nm wide ($l_g$) and are
    separated by $l_{LC}$ = 160 nm and $l_{CR}$ = 220 nm. A double quantum dot (dot 1, D1; dot 2, D2) can be formed
    by applying negative gate voltages to gates GL, GC and GR. A microwave field can be applied to
    the center gate GC.}
    \label{devices}
  \end{center}
\end{figure}

A scanning electron microscope (SEM) image of the first device is
shown in Fig$.$ \ref{devices}(a). Metal gates are deposited on top
of a GaAs/AlGaAs heterostructure with a 2DEG 100 nm below the
surface \cite{Vaart:1995}. Applying a negative voltage to all
gates depletes the 2DEG underneath them and forms two quantum
dots. Current can flow from the large electron reservoir on the
left via the three tunnel barriers induced by the gate pairs 1-F,
2-F, and 3-F to the reservoir on the right. The transmission of
each tunnel barrier can be controlled individually by the voltage
on gates 1, 2 or 3. A single quantum dot can be defined in the
2DEG by applying only a voltage to gates 1, 2, I and F (dot 1) or
to gates 2, 3, II and F (dot 2). In this way, the individual dots
can be characterized and their properties compared to those of the
double
dot.\\
\indent The second device is schematically shown in Fig$.$
\ref{devices}(b). First a channel is defined in the 2DEG by
focused-ion-beam (FIB) or electron-cyclotron-resonance (ECR)
etching of an Al$_{0.3}$Ga$_{0.7}$As/GaAs modulation-doped
heterostructure \cite{Fujisawa:1996}. A double quantum dot can be
formed by applying negative voltages to gates GL, GC and GR.\\
\indent All experiments have been performed in a dilution
refrigerator with a base temperature of 10 mK. The effective
electron temperature in the leads is higher and can vary between
$\sim$40 and $\sim$100 mK. A significant source of heating is the
noise coming from the measurement electronics. The filters, used
to attenuate the noise, have to be effective over a very large
band width. They consist of a distributed RC network, usually a
thin resistive wire going through a conducting medium such as
copper powder or silver epoxy. The filters are installed at low
temperature to minimize the thermal noise of the resistors inside.
The filters are integrated with the sample holder in such a way
that all sample wires are carefully shielded once they are
filtered.
\begin{figure}[htbp]
  \begin{center}
    \centerline{\epsfig{file=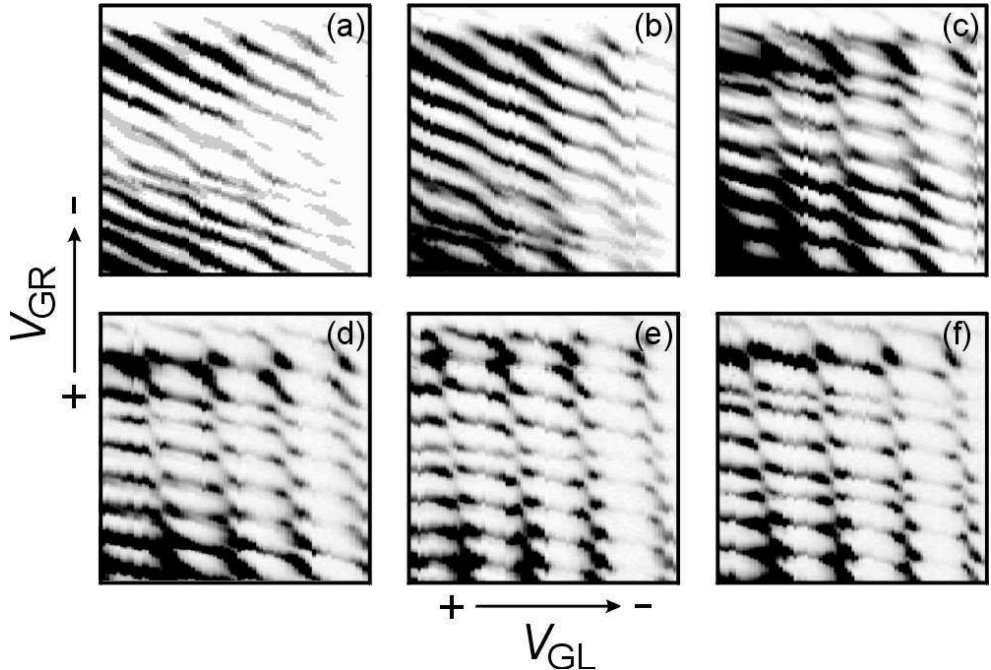, width=13cm, clip=true}}
    \caption{Experimental gray-scale plots of stability diagrams in a device
    similar to that shown in Fig$.$ \ref{devices}(b) for increasingly negative gate
    voltage on the middle gate electrode, GC. Dark (light) gray-scale corresponds to large (small)
    current through the double dot. GL is swept between -500 mV and -530 mV, GR between -840 mV and -900 mV, GC = -660 mV
    (a), -670 mV (b), -690 mV (c), -700 mV (d), -710 mV (e) and
    -720 mV (f).}
    \label{exphoney}
  \end{center}
\end{figure}

To effectively create a double quantum dot, all gate voltages need
to be tuned properly. The cross capacitances between the various
gate electrodes, make it difficult to vary just a single parameter
without affecting the others. The stability diagram is of great
value in setting up and characterizing a double quantum dot.
Figure \ref{exphoney} illustrates the process of the creation of a
double dot in a device similar to the one in Fig$.$
\ref{devices}(b). Starting point is the creation of a single large
dot formed by the outer tunnel barriers, GL and GR. The measured
stability diagram (Fig$.$  \ref{exphoney}(a)) resembles Fig$.$
\ref{honeycombs}(c). The successive stability diagrams are
measured for increasingly negative voltages on the middle gate
electrode, GC, thus reducing the coupling between the dots. It is
clearly seen that going from Fig$.$ \ref{exphoney}(a) to Fig$.$
\ref{exphoney}f, the stability diagram gradually evolves into the
characteristic honeycomb structure. The edges of the honeycomb
cells are visible due to off-resonance current. At the edges of a
honeycomb cell, the electrochemical potential of one of the dots
aligns with its neighboring lead (see Fig$.$ \ref{honeyelch}). By
a process called co-tunneling \cite{Averin:1992}, transport can
still take place via an intermediate virtual state. Co-tunneling
processes are suppressed by increasing the tunnel barriers (i.e.
making the gate voltages more negative), as can be seen from
Fig$.$ \ref{exphoney}.
\begin{figure}[htbp]
  \begin{center}
    \centerline{\epsfig{file=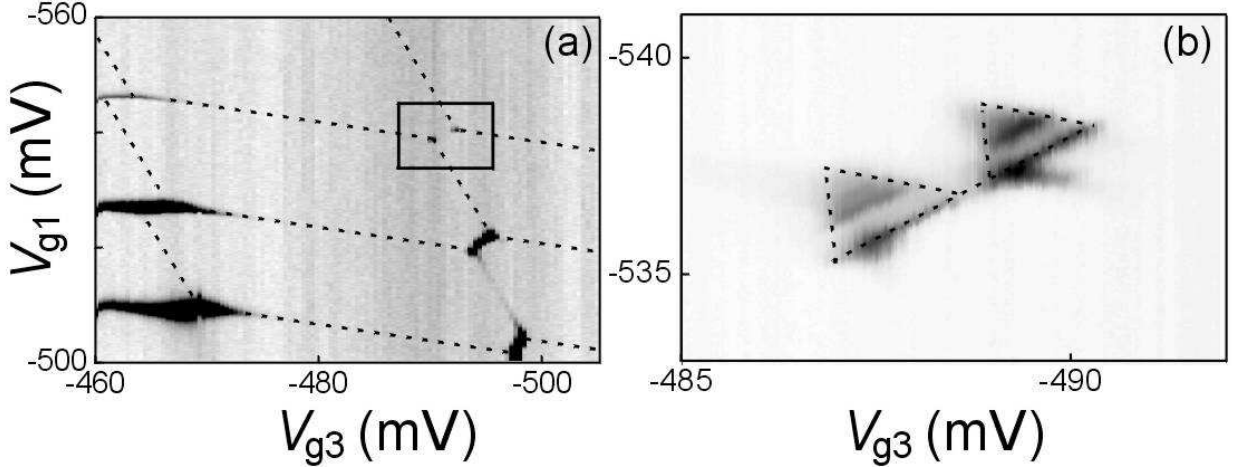, width=17cm, clip=true}}
    \caption{(a) Experimental gray-scale plot of a stability diagram
    in the device of Fig$.$ \ref{devices}(a) at small bias voltage,
    $V$ = 15 $\mu$eV. Dark (light) gray-scale corresponds to large
    (small) current through the double dot. The dashed lines indicate
    the honeycomb cells. (b) Region within the black rectangle of (a)
    at large bias voltage, $V$ = 120 $\mu$eV. The triple points have
    grown into triangles and show clear resonant tunneling lines
    (black stripes), as discussed in section \ref{nonlinquant}. The
    shape of the triangles is accentuated by dashed lines.}
    \label{expfrank}
  \end{center}
\end{figure}

Figure \ref{expfrank}(a) shows a detail of a stability diagram
obtained in the device shown in Fig$.$ \ref{devices}(a). The edges
of the honeycomb cells are indicated by dashed lines. The triple
points within the black square are well separated, whereas the
other ones are still grown together. To separate also those points
outside, the gate voltage on the middle barrier has to be tuned
towards more negative values.

\vspace{0.5cm}

\subsection{Non-linear transport regime}

\label{nonlinear}

\subsubsection{Classical theory}

\begin{figure}[htbp]
  \begin{center}
    \centerline{\epsfig{file=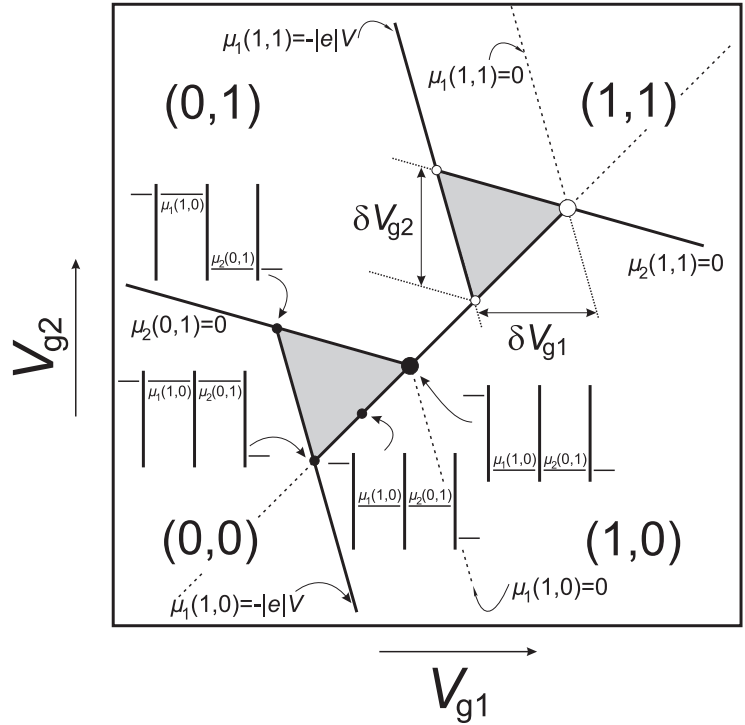, width=9.75cm, clip=true}}
    \caption{Region within the dotted square of Fig$.$ \ref{honeycombs}(a), corresponding to the `unit cell' of
    the double-dot stability diagram, at finite bias voltage. The solid lines separate the charge domains. Classically, the regions of the stability
    diagram where current flows, are given by the gray triangles. In the case of one discrete level per dot,
    as in the schematic pictures, resonant tunneling is only possible along the side of the triangle that
    coincides with the dashed line connecting the original triple-points ({\Large $\bullet$} and {\Large $\circ$}). However, also in this case {\it inelastic
    tunneling} and {\it co-tunneling} still contribute to a finite current within the gray triangles.}
    \label{lbschem}
  \end{center}
  \vspace{-0.25cm}
\end{figure}

We assume that the bias voltage is applied to the left lead
($\mu_L = -|e|V$) and that the right lead is grounded ($\mu_R$ =
0). The bias voltage is coupled to the double dot through the
capacitance of the left lead, $C_L$, and hence also affects the
electrostatic energy of the system. The bias dependence can be
accounted for by replacing $C_{g1(2)}V_{g1(2)}$ with
$C_{g1(2)}V_{g1(2)} + C_{L1(2)}V$ in Eq$.$ (\ref{eqDDenergy}),
where $C_{L1(2)}$ is the capacitance of the left lead to dot 1
(see
Appendix A).\\
\indent The conductance regions at finite bias change from
triple-points to triangularly shaped regions (Fig$.$
\ref{lbschem}). The conditions $-|e|V = \mu_L \geq \mu_1$, $\mu_1
\geq \mu_2$, and $\mu_2 \geq \mu_R$ = 0 determine the boundaries
of the triangular regions. The dimensions of the triangles $\delta
V_{g1}$ and $\delta V_{g2}$ (see Fig$.$ \ref{lbschem}) are related
to the applied bias voltage as follows

\begin{eqnarray}
\alpha _{1}\delta V_{g1}=\frac{C_{g1}}{C_{1}}|e|\delta V_{g1}=|eV| \nonumber \\
\alpha _{2}\delta V_{g1}=\frac{C_{g2}}{C_{2}}|e|\delta V_{g2}=|eV|
\label{alphas}
\end{eqnarray}
\\

\noindent where $\alpha_1$ and $\alpha_2$ are the conversion
factors between gate voltage and energy. Combining Eqs$.$
(\ref{eqDeltaVg}), (\ref{eqDeltaVgm}) and (\ref{alphas}), we can
calculate the values of the total capacitances $C_{1,2}$ and
mutual capacitance $C_m$.

\begin{figure}[htbp]
  \begin{center}
    \centerline{\epsfig{file=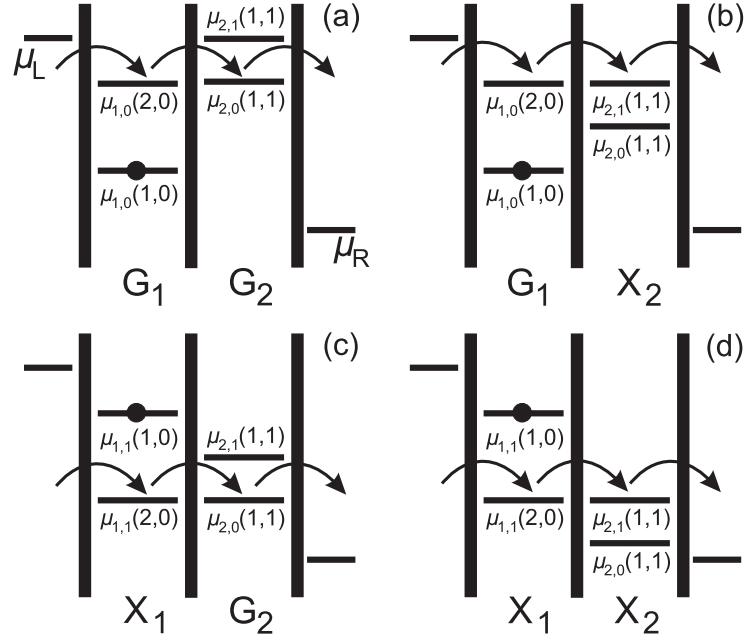, width=9.75cm, clip=true}}
    \caption{Schematic diagrams showing the possible alignments of the electrochemical potentials in the
    case of two levels per dot. (a) The first electrochemical potentials to align correspond to the ground
    states of both dots, G$_1$ and G$_2$. (b) When moving down the levels in the right dot, the next
    states to align are the ground state of the left dot, G$_1$, and the first excited state of the right
    dot, X$_2$. (c) Shifting the levels of the right dot further down, results in transport
    through the first exited state of the left dot, X$_1$ and the right dot ground state, G$_2$.
    (d) Finally, the excited states, X$_1$ and X$_2$ align.}
    \label{non-lin-schematic}
  \end{center}
 \end{figure}

\subsubsection{Quantized states}

\label{nonlinquant}

For sufficiently large bias voltages, multiple discrete energy
levels can enter the bias window. In this case, not only ground
states, but also excited states contribute to the conductance. For
the illustrative case of two levels per dot, the four possible
alignments of the electrochemical potentials are shown in Fig$.$
\ref{non-lin-schematic}. Note that the electrochemical
\begin{figure}[htbp]
  \begin{center}
    \centerline{\epsfig{file=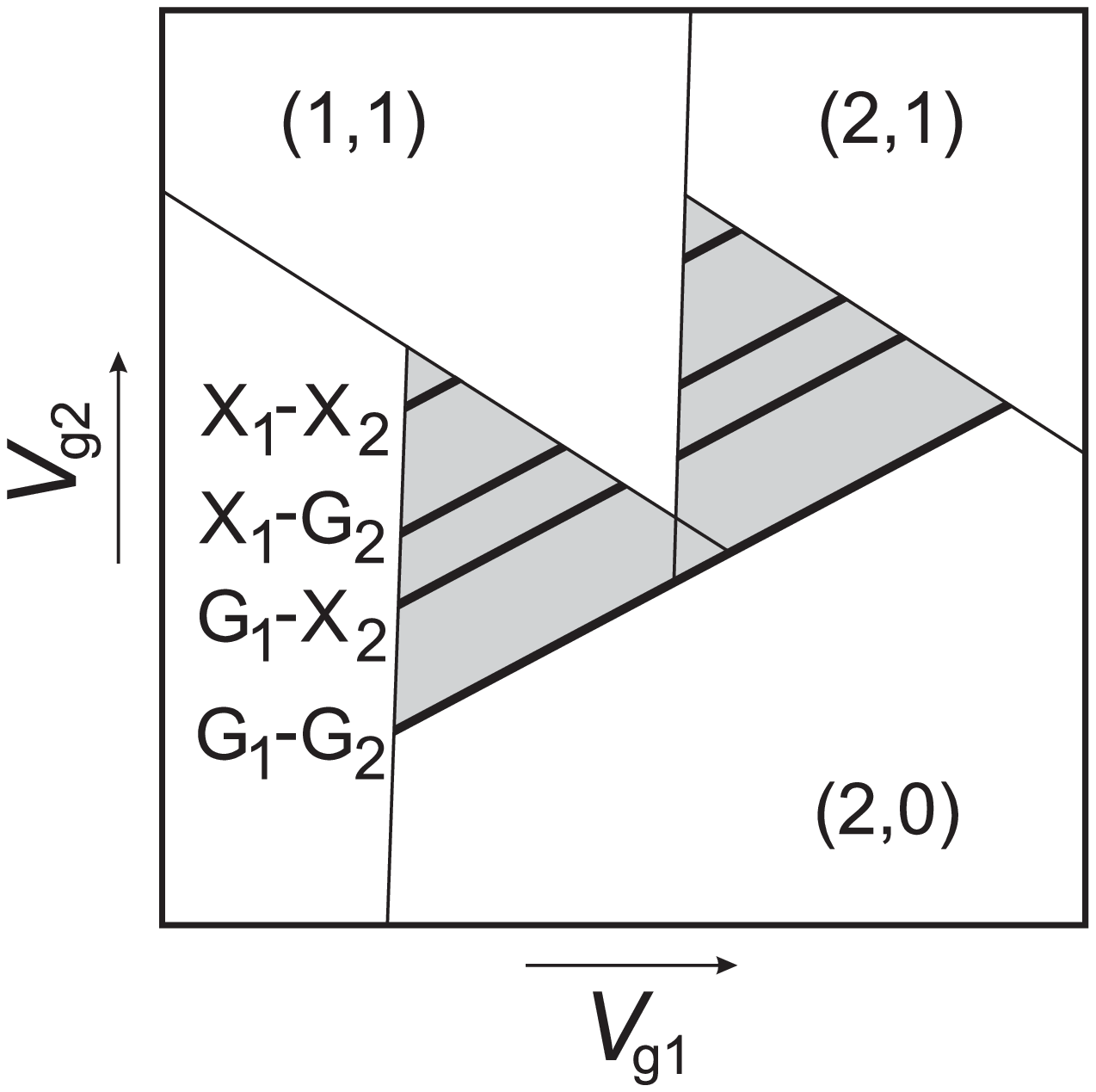, height=7.5cm, clip=true}}
    \caption{Schematic stability diagram corresponding to
    the finite-bias diagrams of Fig$.$ \ref{non-lin-schematic}.
    The black solid lines within the gray triangles correspond,
    from bottom to top, to the level alignments shown in Fig$.$
    \ref{non-lin-schematic}(a)-(d), respectively.}
    \label{qtriangles}
  \end{center}
\end{figure}
potentials are drawn for the situation where one electron is on
the double dot and a second one is tunneling on to it. Due to
Coulomb blockade, not less than one and not more than a total of
two electrons is allowed on the double dot. The labeling of the
electrochemical potentials, using the notation introduced in
section \ref{linquant} is straightforward, except for tunneling
through the excited state of dot 1 in Figs$.$
\ref{non-lin-schematic}(c),d. Although the second electron is
tunneling into the lowest level available in dot 1, this level is
only accessible because dot 1 is in an excited state. For that
reason, we choose the label $\mu_{1,1}(2,0)$ (instead of
$\mu_{1,0}(2,0)$). The successive alignment of ground and excited
states leads to resonances within the conductance triangles, as
shown in Fig$.$ \ref{qtriangles}. The off-resonance conductance in
the grey triangles is due to inelastic processes
\cite{Vaart:1995,Fujisawa:1998} and co-tunneling
\cite{Averin:1992}. Note that $V$ is so large that the two
triangles partly overlap.
\begin{figure}[htbp]
  \begin{center}
    \centerline{\epsfig{file=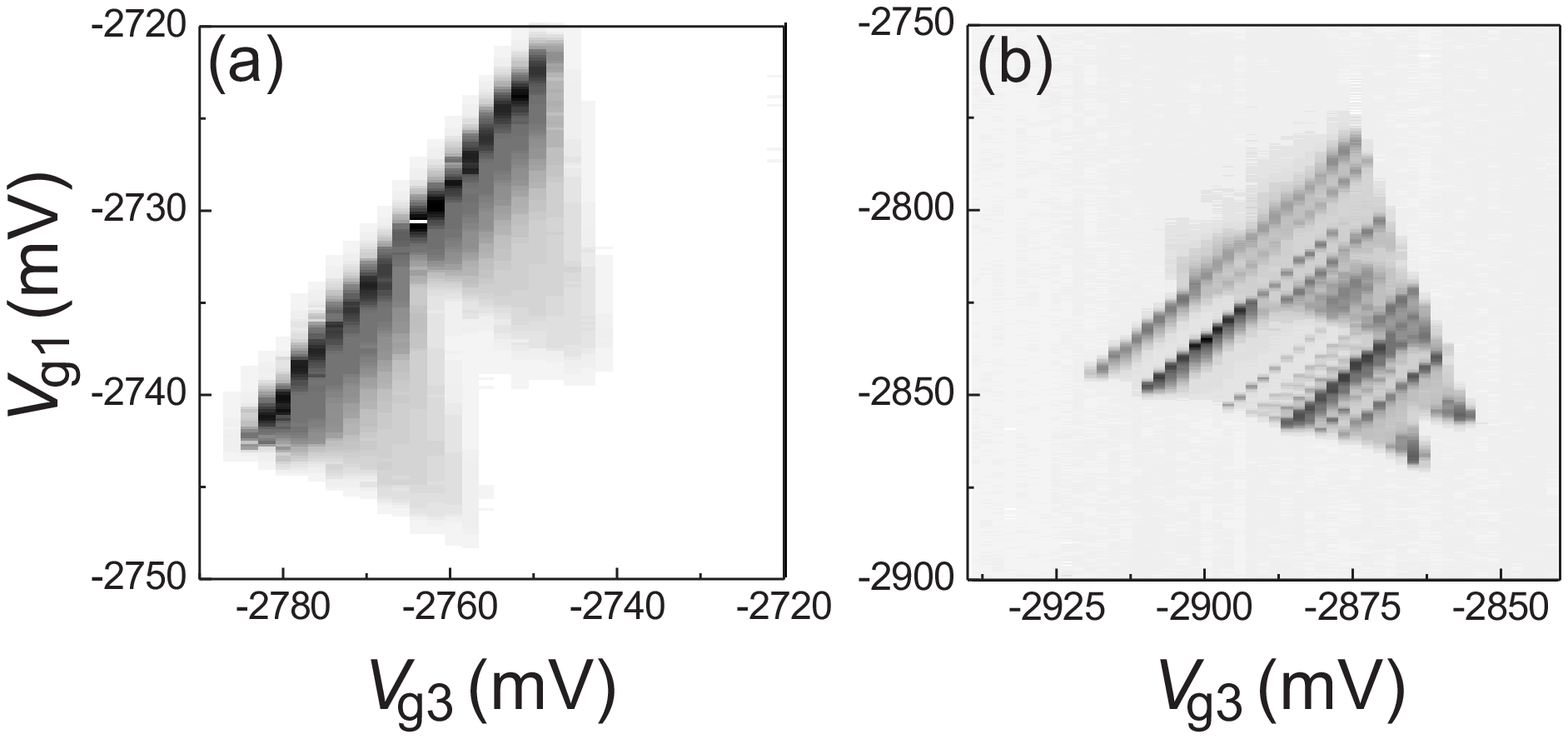, width=12cm, clip=true}}
   \caption{Experimental gray-scale plots of stability diagrams in the non-linear
    regime, obtained in the device of Fig$.$ \ref{devices}(a). Dark (light) gray-scale corresponds to large
    (small) current through the double dot. The bias
    voltage between source and drain contacts is 200 $\mu$V (a) and 1 mV (b).}
    \label{mutriangles}
  \end{center}
\end{figure}

Figure \ref{expfrank}(b) shows the triple points within the black
rectangle of Fig$.$ \ref{expfrank}(a) at finite bias. The
triangular regions are clearly visible as well as the resonances
within the triangles. The growth of the triangular regions with
increasing bias voltage is illustrated in Fig$.$
\ref{mutriangles}. Whereas only the ground state resonance is
observed in Fig$.$ \ref{mutriangles}(a), multiple resonances
appear within the triangles of Fig$.$ \ref{mutriangles}(b).

\section{Resonant tunneling}

\label{restun}

\begin{figure}[htbp]
  \begin{center}
    \centerline{\epsfig{file=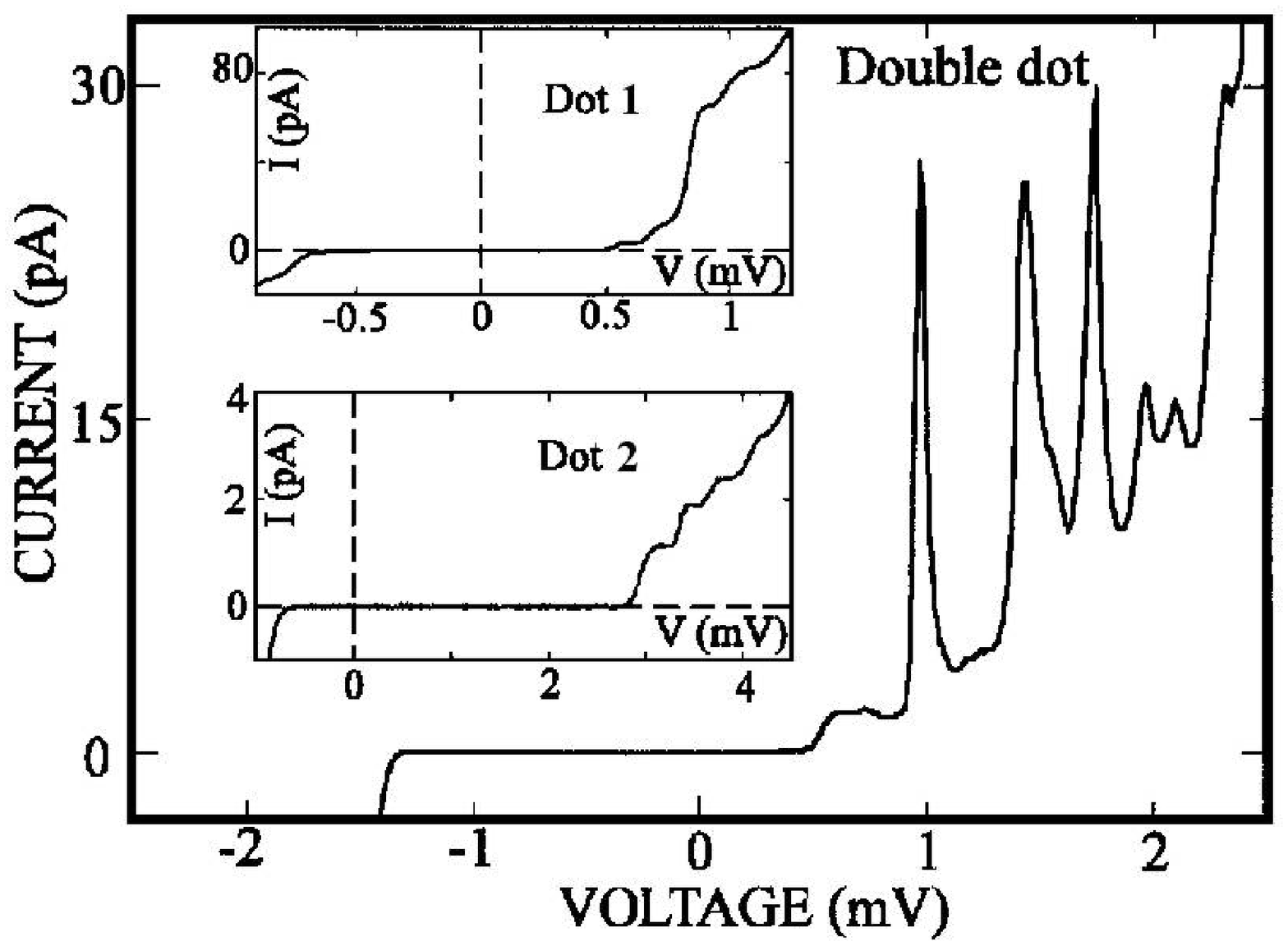, width=10cm, clip=true}}
    \caption{$I$-$V$ curve of the double dot, showing sharp
    resonances in the current when two discrete levels align.
    Upper inset: $I$-$V$ curve of dot 1. Lower inset: $I$-$V$
    curve of dot 2. Both insets show a suppression of the
    current at low voltages due to the Coulomb blockade and a
    stepwise increase of the current due to the discrete energy
    spectrum of the dot (from \cite{Vaart:1995}).}
    \label{nijs2}
  \end{center}
  \vspace{-0.5cm}
\end{figure}

In this section we discuss resonant tunneling experiments through
the double dot of Fig$.$ \ref{devices}(a) with discrete energy
levels \cite{Vaart:1995}. We show that, under appropriate
conditions, the resonance widths are only determined by the
lifetime of the discrete energy states, independent of the
electron temperature in the leads. A small asymmetric deviation
from the Lorentzian
resonance shape is attributed to inelastic tunnel processes.\\
\indent The current-voltage ($I$-$V$) curves of the single quantum
dots in Fig$.$ \ref{nijs2} provide two clear signatures for the
presence of both Coulomb blockade effects and discrete levels. At
low bias voltages, the current through the dot is suppressed by
the Coulomb blockade \cite{Kouwenhoven:1997}. Increasing the bias
voltage lifts the blockade. The current shows a stepwise increase:
each time when an additional level enters the bias window $-|e|V$,
an extra transport channel is opened and the current increases
\cite{Su:1992,Gueret:1992,Johnson:1992,Foxman:1993}. Hence, the
voltage spacing of the current steps directly reflects the energy
spacing of the levels. For the average level spacing, $ \delta $,
we obtain $ \delta_1 $ = 125 $\mu$eV for dot 1 (upper inset) and
$\delta_{2}$ = 225 $\mu$eV for dot 2 (lower inset). The difference
in these two energies reflects the different lithographic sizes of
the two dots (see Fig$.$ \ref{devices}(a)). Accounting for the
depletion areas, we estimate that dot 1 has an effective diameter
of 240 nm and contains about $N_1$ = 90 electrons, while dot 2 has
an effective diameter of 200 nm and contains roughly $N_{2}$ = 60
electrons. Using the Fermi energy, $E_F$, at bulk density, we
estimate $ \delta_1 \approx 2 E_F/N_1$ = 150 $\mu$eV and $
\delta_{2} \approx 2 E_F/N_{2}$ = 230 $\mu$eV. This is in good
agreement with the estimates obtained from the $I$-$V$ curves.
From the dimensions of the Coulomb diamonds
\cite{Kouwenhoven:1997,Pothier:1992} we obtain the charging
energies $E_C$ for adding an electron to the dot : $E_{C1}$ = 1.1
meV (dot 1) and $E_{C2}$ = 1.8 meV (dot 2).

\begin{figure}[h]
  \begin{center}
    \centerline{\epsfig{file=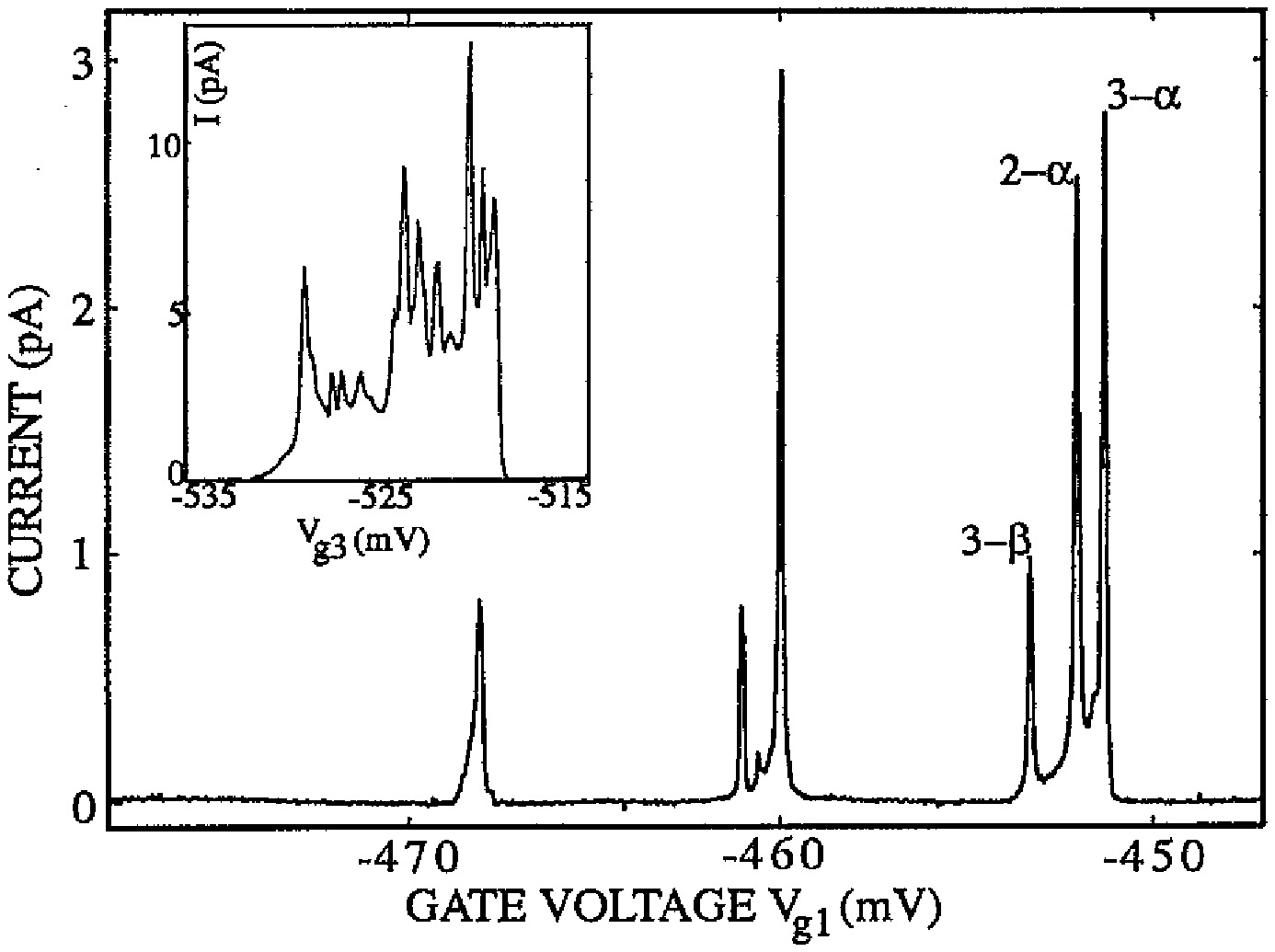, width=10cm, clip=true}}
    \caption{Current through the double dot versus gate voltage $V_{g1}$ using a bias voltage $V$ = 280 $\mu$V.
    Inset: The current through the double dot as a function of $V_{g3}$ with $V$ = 1 mV, showing that the number of
    resonances increases with bias voltage (from \cite{Vaart:1995}).}
    \label{nijs3}
  \end{center}
  \vspace{-0.5cm}
\end{figure}

We focus on the role of the discrete levels and consider the
charging energies as constant offsets in the transport conditions.
Figure \ref{nijs2} shows an $I$-$V$ curve of the double dot with
all three tunnel barriers set in the weak-tunneling regime. The
Coulomb blockade suppresses the current through the double dot at
low bias voltages. At larger bias the current shows sharp
resonances. The spacing of the resonances is about 250 $\mu$eV.
This is of the same order as the level spacing in the single
dots.\\
\indent The same resonances are seen when we sweep the gate
voltage. Figure \ref{nijs3} shows the current through the double
dot versus the gate voltage on gate 1, $V_{g1}$, with $V$ = 280
$\mu$eV. This corresponds to a vertical cut through a stability
diagram as shown in Fig$.$ \ref{mutriangles}. The current shows
three groups of sharp resonances separated by regions of zero
current with a period $\Delta V_{g1}$ = 9 mV in gate voltage
$V_{g1}$. With only dot 1 formed, we observe Coulomb oscillations
as a function of $V_{g1}$ with the same period $\Delta V_{g1}$;
each period thus corresponds to a change of one electron in dot 1,
while keeping the number of electrons on dot 2, $N_{2}$, constant.
$\Delta V_{g1}$ corresponds to the horizontal dimension of the
honeycomb unit cell, as indicated in Fig$.$ \ref{spacings} in
section \ref{linclass}.

\begin{figure}[htbp]
  \begin{center}
    \centerline{\epsfig{file=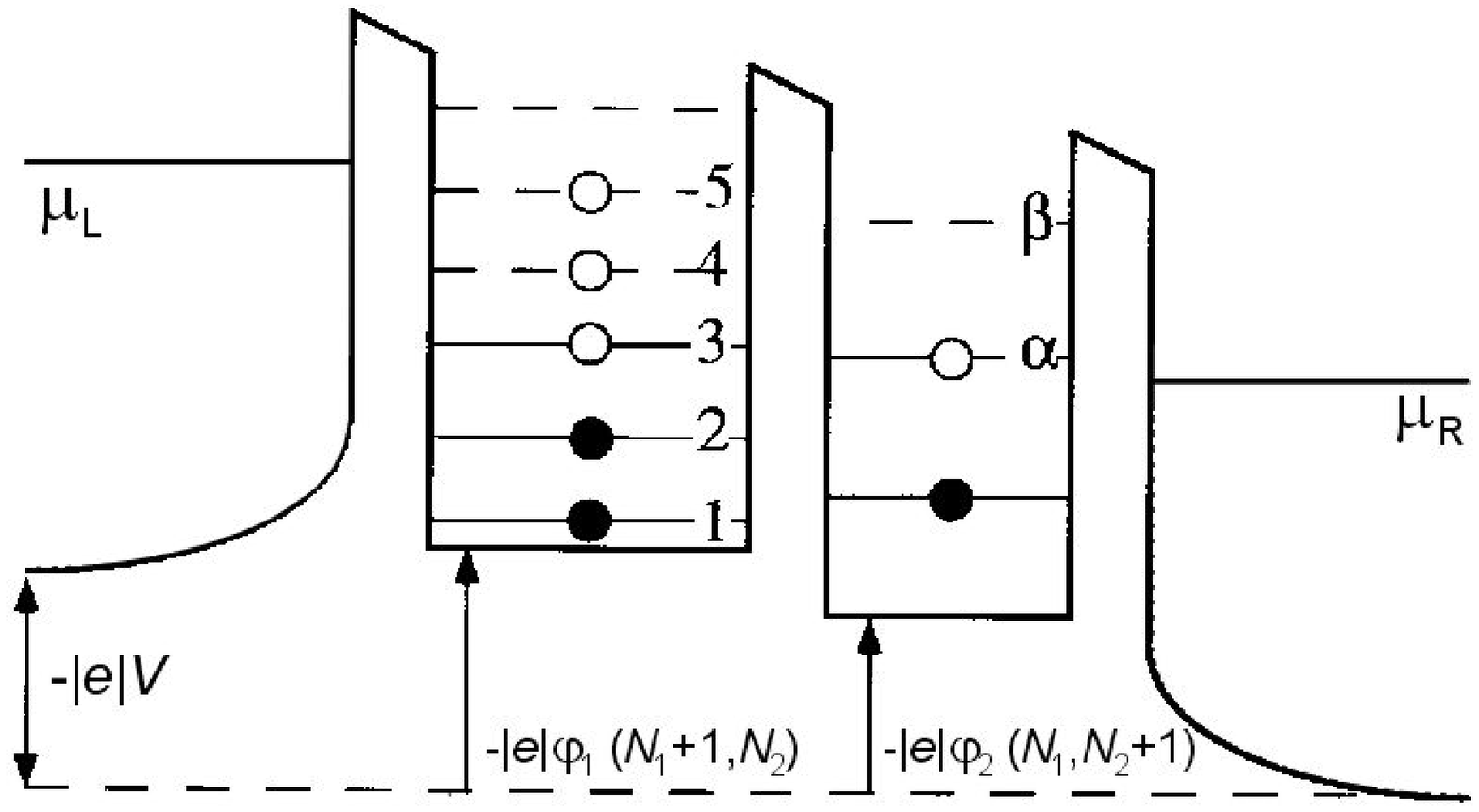, width=10cm, clip=true}}
    \caption{Schematic potential landscape of the double quantum
    dot, where $\mu_L$ and $\mu_R$ denote the
    electrochemical potentials of the left and right reservoirs
    and $V$ the bias voltage across the double dot. The 0D states
    in dot 1 are denoted by levels 1 to 5 and in dot 2 by levels
    $\alpha$ and $\beta$ (from \cite{Vaart:1995}).}
    \label{nijs1b}
  \end{center}
  \vspace{-0.5cm}
\end{figure}

When elastic tunnel processes are the dominant transport
mechanism, the current through the double dot is resonantly
enhanced only when two levels in dot 1 and 2 align, as explained
in section \ref{nonlinquant}. Tuning the level alignment with $V$
or $V_{g1}$ gives rise to the sharp resonances in Figs$.$
\ref{nijs2} and \ref{nijs3}. Resonant tunneling through the double
dot is illustrated in the schematic potential landscape of the
double dot in Fig$.$ \ref{nijs1b}. This figure shows a few of the
levels in dot 1 (levels 1 to 5) and dot 2 (levels $\alpha$ and
$\beta$). The electrostatic potentials $\varphi_1$ and
$\varphi_{2}$ are tuned in such a way that transport through the
double dot is possible only via the charge states $(N_1, N_{2})
\rightarrow (N_1 + 1, N_{2}) \rightarrow (N_1, N_{2} + 1)
\rightarrow (N_1, N_{2})$. The finite bias voltage gives an
electron from the left reservoir three choices to tunnel into dot
1: it can tunnel to one of the unoccupied levels 3, 4 or 5. This
changes the electrostatic potential $\varphi_1$ by the charging
energy $E_{C1}$ (the levels are drawn at the positions applicable
after an electron has occupied one of them). When dot 1 relaxes to
the ground state (the incoming electron occupying level 3), the
electron can tunnel via level $\alpha$ to the right reservoir.

Note that if dot 1 does not immediately relax to the ground state,
but remains in an excited state, with an electron occupying either
level 4 or 5, electron transport through the double dot is
temporarily blocked. The electron is `trapped' within dot 1. Only
after relaxation to the ground state a next tunnel event can
occur. If the relaxation rate is small on the scale of the
tunneling rates through the barriers, the inclusion of levels 4
and 5 within the bias window could therefore lead to a {\it
decrease} of the current through the double dot. On the other
hand, in case of fast relaxation, the enhanced tunnel probability
when also the levels 4 and 5 lie within the bias window, could
lead to an increase in the current. Note that next to intra-dot
relaxation, also inelastic tunneling from level 4 or 5 to either
level $\alpha$ or $\beta$ can occur. This process is accompanied
by emission of a boson (usually phonons \cite{Fujisawa:1998}) and
contributes to the off-resonance current in Fig$.$ \ref{nijs3}.

The resonances in a particular group in Fig$.$ \ref{nijs3} can be
identified with the energy diagram of Fig$.$ \ref{nijs1b}. The
first resonance occurs when level 3 aligns with level $\alpha$
(peak 3-$\alpha$). This corresponds to the rightmost peak in
Fig$.$ \ref{nijs3}. Increasing $-|e| \varphi_1$ by making $V_{g1}$
more negative, brings transport off-resonance until level 2 aligns
with $\alpha$ (peak 2-$\alpha$) followed by the third peak
3-$\beta$. Continuing to sweep $V_{g1}$ increases the energy of
level 3 above the electrochemical potential $\mu_L$ of the left
reservoir. This blocks transport and removes an electron from dot
1 permanently. The next group of resonances is observed when
$V_{g1}$ is changed by one Coulomb oscillation period $\Delta
V_{g1}$ (see Fig$.$ \ref{nijs3}). Note that the number of
resonances decreases in the next two groups. Sweeping $V_{g1}$
also shifts the levels in dot 2, due to a small cross-capacitance
between gate 1 and 2. Transport is possible until level $\alpha$
is shifted above $\mu_L$ ($V_{g1}$ $<$ -470 mV).\\
\indent The level spacing is obtained by converting gate voltage
to energy \cite{Grabert:1992} . This yields an energy separation
of resonances 2-$\alpha$ and 3-$\alpha$ by 70 $\mu$eV, which is
the energy separation of levels 2 and 3. In the same way we find
for levels $\alpha$ and $\beta$ a separation of 200 $\mu$eV. Both
values are in good agreement with the typical values we found
above. On increasing $V$, we observe that the number of resonances
in a particular group increases. The inset to Fig$.$ \ref{nijs3}
shows approximately 11 resonances as $V_{g3}$ is swept. These
observations are in agreement with the resonant tunneling picture
of Fig$.$ \ref{nijs1b}: when $V$ is larger, more levels can align.
\begin{figure}[htbp]
  \begin{center}
    \centerline{\epsfig{file=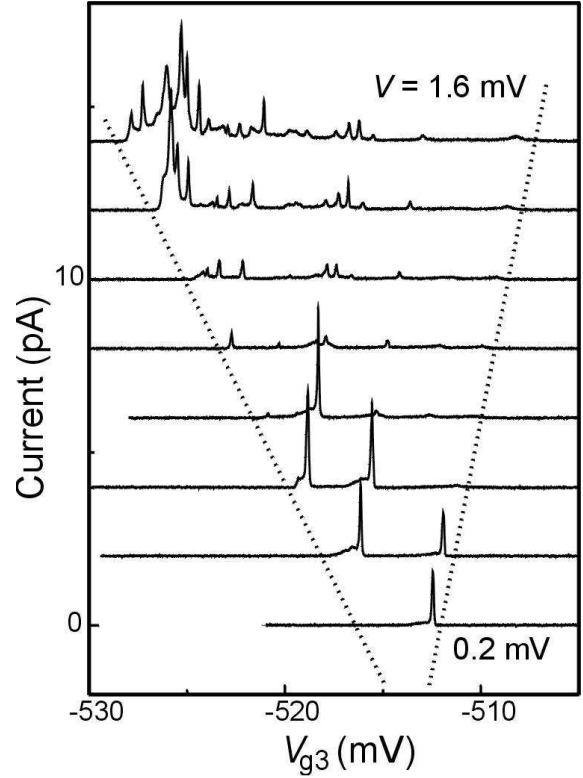, width=7.5cm, clip=true}}
    \caption{Current through the double dot of Fig$.$ \ref{devices}(a)
    versus the voltage on gate 3, $V_{g3}$ for bias voltages, $V$,
    between 0.2 mV (lower trace) and 1.6 mV (upper trace). The
    traces have been given an offset proportional to their bias
    voltage for clarity.}
    \label{waaier}
  \end{center}
  \vspace{-0.5cm}
\end{figure}

Generally, the relaxation rate to the ground state is not
necessarily higher than the tunnel rate through the dot. In Fig$.$
\ref{waaier} the amplitude of the ground state resonance (see
lower curve) clearly decreases with increasing bias voltage. At
the same time new resonances appear, having a larger amplitude
than the ground state resonance. This implies that in general
transport through excited states can play a significant role.

\begin{figure}[htbp]
  \begin{center}
    \centerline{\epsfig{file=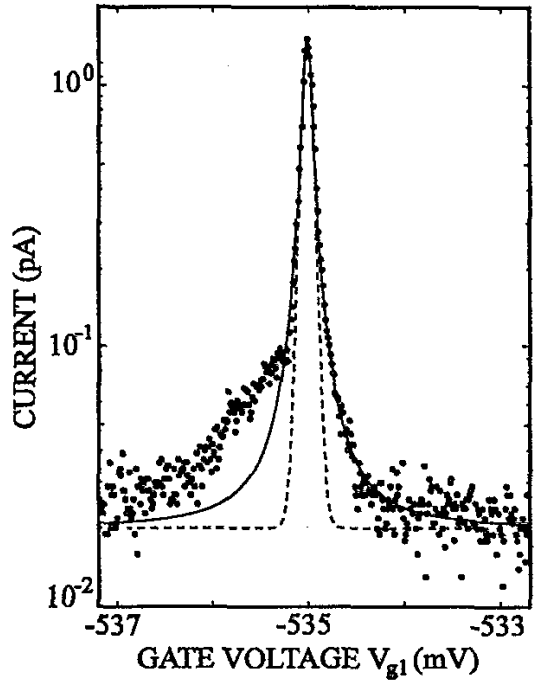, width=7cm, clip=true}}
    \caption{Enlarged resonance measured in a second device of identical design using a bias voltage of 400 $\mu$eV.
    The data points (black dots) are fit to a Lorentzian line shape (solid line). For comparison we plot a thermally
    broadened resonance with a fitted temperature $T$ = 34 mK (dashed line) (from \cite{Vaart:1995}).}
    \label{nijs4}
  \end{center}
  \vspace{-0.5cm}
\end{figure}

When a discrete level in dot 1 is at a distance much larger than
the thermal energy, $k_B T$, from the electrochemical potential of
the left lead, dot 1 acts as low-temperature-pass filter for dot
2, such that only cold electrons contribute to the current
\cite{Kouwenhoven:1995}. If the energy levels in the dots are
separated by more than $k_B T$, the occupation of excited states
becomes suppressed. This effectively leaves the dot at zero
temperature. Hence, the double dot geometry allows for an accurate
measurement of the intrinsic line width of the discrete levels,
which is not averaged by the Fermi-Dirac distribution of the
electrons in the reservoirs. In other words, an energy resolution
better than $k_B T $ can be obtained. The line shape of a
resonance is lorentzian when only elastic tunneling is important
\cite{Nazarov:1993,Stoof:1997}. In our geometry (assuming a bias
voltage sufficiently large such that electrons must tunnel from
the left lead to the left dot to enter the system, and must tunnel
from the right dot to the right lead to leave it again) the
current is given by

\begin{equation}
I(\Delta E)=e\frac{\Gamma _{3}|t_{12}|^{2}}{(\Delta
E/h)^{2}+\frac{\Gamma _{3}^{2}}{4}+|t_{12}|^{2}\left(
2+\frac{\Gamma _{3}}{\Gamma _{1}}\right)} \label{lorentzian}
\end{equation}
\\
\noindent where $\Delta E$ is the energy difference between two
discrete energy levels in the two dots, $\Gamma_1$ is the tunnel
rate from the left lead to dot 1, $|t_{12}|$ is the modulus of the
tunnel coupling between the two dots and $\Gamma_3$ is the tunnel
rate from dot 2 to the right lead. Note that for elastic tunneling
the resonance width is only determined by the lifetime of the 0D
states and independent of temperature.\\
\indent Figure \ref{nijs4} shows a single resonance (black dots).
The right-hand side of the peak fits very well with the lorentzian
line shape of Eq$.$ (\ref{lorentzian}) (solid line), while the
left-hand side shows a deviation from the lorentzian fit. The only
free fit parameter is the full width at half maximum, FWHM = 5
$\mu$eV. From the maximum current and the width of the resonance
we find with Eq$.$ (\ref{lorentzian}) a tunnel coupling $|t_{12}|$
$\approx$ 0.2 $\mu$eV and a tunneling rate between the right dot
and the right lead $\Gamma_3 \approx 10 \mu$eV. For comparison, we
have fitted the resonance with a thermally broadened resonance
$I(\Delta E) \sim$ cosh($\Delta E/2kT)^{-2}$ (dashed line)
\cite{Beenakker:1991}. The top is fit very well for $T$ = 34 mK,
but there is a large deviation in both tails of the resonance. On
the right-hand side, the deviation can be accounted for by the
lorentzian broadening. At the left-hand side, the deviation
consists of two components. The first one is the same lorentzian
broadening as observed on the right-hand side. The second one,
however, is an asymmetric contribution only occurring on this side
of the resonance. The asymmetric contribution to the current
appears at the side where an electron tunnels from a higher to a
lower electrochemical potential. Upon reversing the sign of $V$,
we find that the asymmetry appears at the other side of the
resonance. As shown in Ref$.$ \cite{Fujisawa:1998}, this is due to
inelastic tunnel processes. In such a process, an electron can
tunnel inelastically and spontaneously emit its energy as a photon
or a phonon.

\section{Magnetic field spectroscopy}

\label{magnetiz}

In this section we measure the energy evolution versus magnetic
field, $B$, of energy states near the Fermi energy, $E_F$, in the
double quantum dot shown in Fig$.$ \ref{devices}(a). As a function
of $B$ and the voltage on gate 3, $V_{g3}$, we observe crossings
and anti-crossings between Coulomb peaks. The resolution is high
enough that avoided crossings in the spectrum of a quantum dot can
be resolved \cite{Oosterkamp:1998a}. To our knowledge, these are
the only existing data revealing intra-dot level repulsion in a
quantum
dot system.\\
\indent The experiments are performed in the weak coupling limit,
such that mixing between quantum states in one dot with states in
the other dot or in the leads is negligible (see section
\ref{spectro}). We sweep the gate voltages over
\begin{figure}[htbp]
  \begin{center}
    \centerline{\epsfig{file=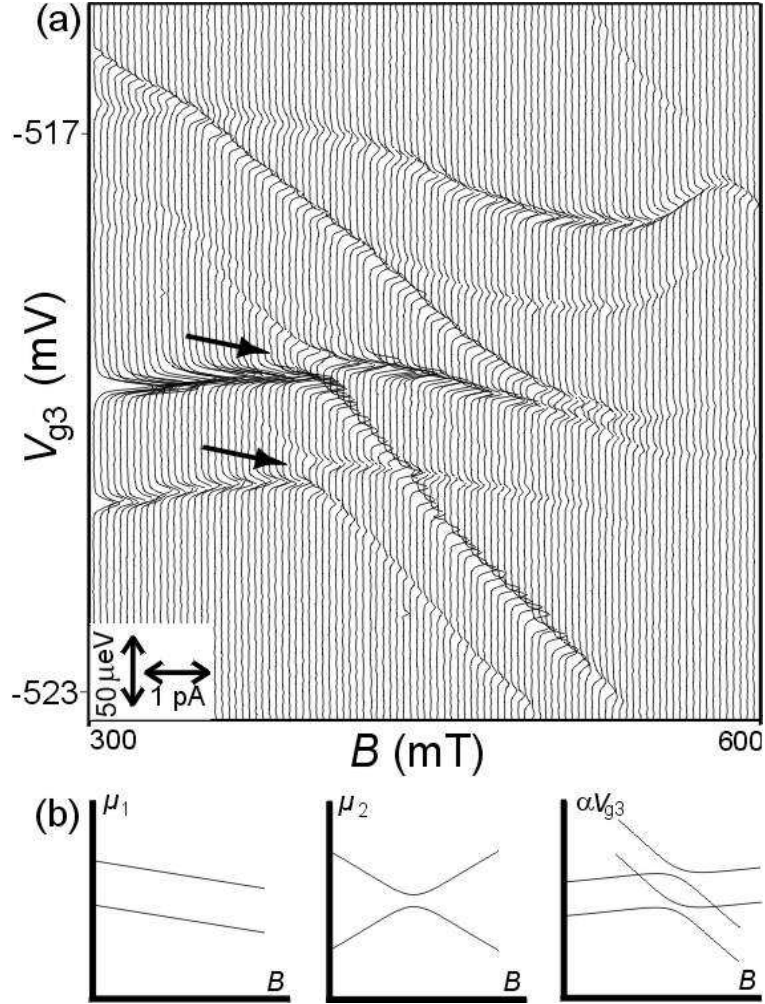, width=10cm, clip=true}}
    \caption{(a) Current through the double dot sweeping the voltage on gate 3, $V_{g3}$, at
    different magnetic fields, $B$. The curves have been given an offset for clarity.
    From the leftmost to the rightmost curve, $B$
    increases from 300 mT to 600 mT in 3 mT increments. The conversion
    factor $\alpha$ between $V_{g3}$ and the electrostatic potential of dot 2, $\alpha$
    = -63 $\mu$eV/mV, is determined through independent
    measurements from which we deduced the energy scale, indicated by the
    vertical arrow in the lower left corner ($\alpha$ does not change in this
    magnetic field range). (b) The
    first two diagrams show how levels may evolve in each of the
    two dots as a function of $B$. When these four
    levels are scanned along each other by sweeping $V_{g3}$, this
    results in peak positions as sketched in the rightmost diagram.}
    \label{Bspectrofig}
  \end{center}
  \vspace{-0.5cm}
\end{figure}
small ranges and focus on a particular charging transition;
i$.$c$.$ transitions between ($N_1$+1,$N_2$) and ($N_1$,$ N_2 $+1)
only. Since we discuss only one transition at a time, we can, for
simplicity, leave out the Coulomb energies from the discussion and
concentrate on the alignment of discrete energy
levels.\\
\indent Using the notation introduced in section \ref{linquant},
we label the accompanying electrochemical potentials
$\mu_{1,n}(N_1,N_2)$ for dot 1 and $\mu_{2,n}(N_1,N_2)$ for dot 2
(or simply $\mu_{1(2),n}$). The condition for tunneling between
the lowest possible states, i$.$e$.$ from ground state to ground
state, is $\mu_{1,0}(N_1 +1 ,N_2)$ = $\mu_{2,0}(N_1,N_2+1)$. To
tunnel from the first excited state of dot 1 to the ground state
of dot 2, the condition becomes $\mu_{1,1}(N_1 +1 ,N_2)$ =
$\mu_{2,0}(N_1,N_2+1)$. The changes in the electrochemical
potential $\mu_{1,n+1} - \mu_{1,n}$ and $\mu_{2,n+1} - \mu_{2,n}$
are typically $\sim$ 150-200 $\mu$eV.\\
\indent Figure \ref{Bspectrofig}(a) shows a typical set of current
traces for different magnetic fields while sweeping $V_{g3}$. The
bias voltage $V$ = 1.2 mV is such that several discrete levels in
each dot are between the electrochemical potentials of the two
leads. This is similar as in the inset to Fig$.$ \ref{nijs3}, but
now measured for different magnetic fields. The {\it change} in
Coulomb peak position versus $B$, $\Delta V_{g3}^{peak}(\Delta
B)$, is proportional to the difference in the $B$-evolution of the
electrochemical potentials $\mu_{1,n}(B)$ and $\mu_{2,n}(B)$
\begin{align}
\Delta \mu(\Delta B) = & \hspace{0.1cm} [\mu_{1,n}(B+\Delta B) -
\mu_{1,n}(B)] - [\mu_{2,n}(B+\Delta B) - \mu_{2,n}(B)] \nonumber \\
= & \hspace{0.1cm} \alpha \Delta V_{g3}^{peak}(\Delta B)
\end{align}
where $\alpha$ is the conversion factor between $V_{g3}$ and the
electrostatic potential of dot 2. Note that if the states
$\mu_{1,n}(B)$ and $\mu_{2,n}(B)$ have the same $B$-dependence,
the Coulomb peak position does not change. The energy resolution
of $\Delta \mu(\Delta B)$ is $\sim$ 5 $\mu$eV, corresponding to
$k_B T \sim$ 50 mK.\\
\indent The data in Fig$.$ \ref{Bspectrofig}(a) contain several
interesting features. First, we observe crossings between
different peaks as well as anti-crossings (two are indicated by
arrows). Second, pairs of peaks exhibit the same $B$-dependence.
These are general features that we observe at several charge
transitions (that is for several choices of $N_1$,$N_2$).
Independent measurements on {\it one} of the individual dots also
show states evolving in pairs below $B \sim 0.5$ T. The observed
pairing and (anti-)crossing of the Coulomb peaks in the double dot
experiments can then be explained as shown schematically in Fig$.$
\ref{Bspectrofig}(b). Suppose two levels in one dot have an
anti-crossing in their $B$-dependence. Then two paired levels in
the other dot, having the same $B$-dependence, both probe this
anti-crossing. At the points where two Coulomb peaks actually
cross, two levels in dot 1 align with two levels in dot 2
simultaneously (though only one electron can
tunnel at a time, due to Coulomb blockade).\\
\indent For the interpretation of the data as schematically given
in Fig$.$ \ref{Bspectrofig}(b), tunneling through the excited
state of dot 1 is a key ingredient. If dot 1 would relax to its
ground state much faster than the tunnel rate through the barriers
after an electron has tunneled onto it via $\mu_{1,1}$, we would
only observe the two lower traces in the rightmost diagram of
Fig$.$ \ref{Bspectrofig}(b). However, our data suggest a
relatively slow relaxation rate between the excited and ground
state of dot 1. Recent experiments on (single) quantum dots have
shown that indeed relaxation times can be of the order of $\mu$s
or longer when relaxation to the ground state involves electron
spin flips \cite{Fujisawa:2001}. The condition for the relaxation
rate from the excited state to the ground state in dot 2 is more
subtle. The electron can tunnel onto dot 2 via $\mu_{2,1}$ and
leave dot 2 either directly or after a relaxation process from
$\mu_{2,1}$ to $\mu_{2,0}$ has occurred (this second possibility
requires $\mu_{2,0} \geq \mu_R$). For transport through the double
dot, the relaxation rate in dot 2 does not necessarily need to be
slow as well. However, if the relaxation rate would be too high,
the anti-crossing as shown in the middle diagram of Fig$.$
\ref{Bspectrofig}(b) would smear out. The constant level spacing
$\mu_{1,1}- \mu_{1,0}$ in dot 1 (see left diagram of Fig$.$
\ref{Bspectrofig}(b)) could be explained by an exchange energy,
e.g. when the upper level would correspond to a spin singlet state
and the lower level to a spin triplet \cite{Tarucha:2000}.

\section{Microwave spectroscopy}

\label{spectro}

In this section we present microwave (0-50 GHz) spectroscopy
experiments
\cite{Kouwenhoven:1997,Fujisawa:1997a,Fujisawa:1997b,Oosterkamp:1997}
on double quantum dots for different coupling and microwave power
regimes \cite{Oosterkamp:1998b,Wiel:1999}. We use photon assisted
tunneling (PAT) processes, as described in sections \ref{weakcPAT}
and \ref{strongcPAT}, to measure the energy differences between
states in the two dots of the devices shown in Figs$.$
\ref{devices}(a) and \ref{devices}(b). Depending on the strength
of the inter-dot coupling, the two dots can form ionic-like
\cite{Vaart:1995,Blick:1996,Livermore:1996,Fujisawa:1997a,Fujisawa:1997b}
or covalent-like bonds
\cite{Blick:1998,Oosterkamp:1998b,Wiel:1999}. In the first case,
the electrons are localized on individual dots, while in the
second case, the electrons are delocalized over both dots. The
covalent binding leads to a symmetric and anti-symmetric state,
whose energy difference is proportional to the tunneling
strength between the dots.\\
\indent For the microwave experiments we make use of a coaxial
cable. From room temperature to the 1K-pot, a 0.085 inch
semi-rigid Be-Cu (inner and outer conductor) coaxial cable is
used. From the 1K-pot to the mixing chamber, we use a 0.085 inch
semi-rigid stainless steel (inner and outer conductor) coax. From
the mixing chamber to the sample, various types of low attenuation
semi-rigid or flexible coaxial cable can be used, since here the
thermal conductivity is no longer a constraint. Finally, the
coaxial cable is capacitively coupled (typically through a 10 pF
capacitor) to one of the gate electrodes of the sample, usually
the center gate. This gate is capacitively coupled to both dots,
and hence part of the incident power can generate a microwave
oscillating potential across the center barrier.

\subsection{Two-level systems}

\label{2level}

\begin{figure}[htbp]
  \begin{center}
    \centerline{\epsfig{file=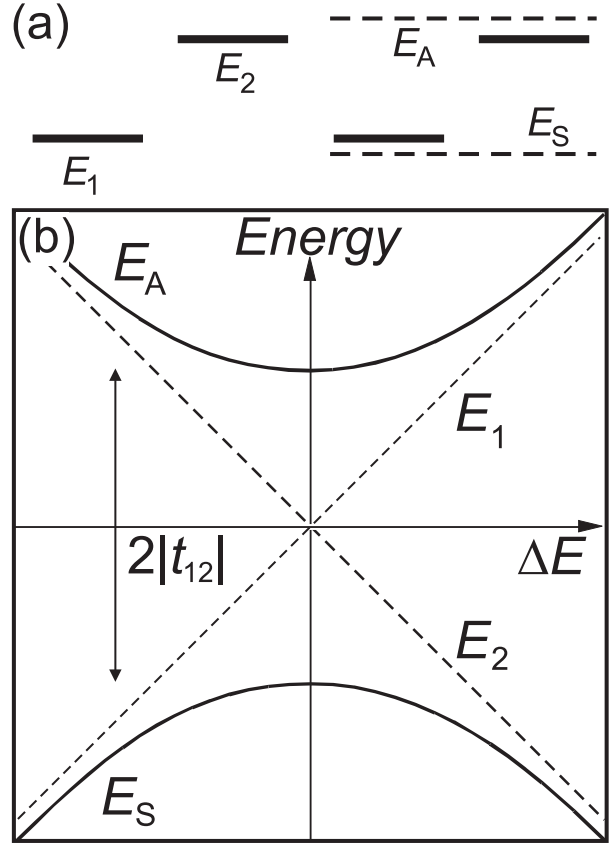, width=8cm, clip=true}}
    \caption{Schematic diagrams of a two-level system. (a) Unperturbed
    energy levels $E_1$ and $E_2$ (solid lines), and energy levels
    belonging to the symmetric state, $E_S$, and the anti-symmetric state, $E_A$.
    (b) Energies $E_S$ and $E_A$ versus the energy
    difference $\Delta E = E_1 -E_2$. For vanishing coupling ($|t_{12}|
    \approx 0$), the levels cross at the origin (dashed straight lines).
    For non-zero coupling, an `anti-crossing' occurs: the curves belonging to $E_S$ and $E_A$
    as function of $\Delta E$ are branches of a hyperbola (solid lines)
    whose asymptotes are the unperturbed levels (see also \cite{Cohen:1977}).}
    \label{2levelfig}
  \end{center}
  \vspace{-0.5cm}
\end{figure}

So far, we assumed a purely electrostatic coupling between both
dots, whereas tunnel coupling was neglected. However, when
electrons can tunnel coherently from one dot to the other at
appreciable rates, the eigenstates become delocalized, extending
over the entire double dot system. In principle, these are
quantum-mechanical many-body states of the two coupled dots. It is
very difficult to give a full description of such a many-body
system. Therefore, we discuss here the elementary case of a
quantum-mechanical two-level system, which is quite adequate in
grasping the physics of a tunnel-coupled double dot. Basically, we
only take into account the topmost occupied level in each dot and
neglect the interaction with electrons in lower energy levels.\\
\indent We consider a double dot consisting of two well-separated
dots, described by a total Hamiltonian $\mathbf{H_0}$
\cite{Cohen:1977}, with eigenstates $\left| \phi_1 \right\rangle$
and $\left| \phi _2\right\rangle$, and eigenenergies $E_1$ and
$E_2$ (Fig$.$ \ref{2levelfig}(a))

\begin{align}
\mathbf{H_{0}}\left| \phi _{1}\right\rangle & = E_{1}\left| \phi _{1}\right\rangle \nonumber \\
\mathbf{H_{0}}\left| \phi _{2}\right\rangle & = E_{2}\left| \phi
_{2}\right\rangle \label{eqH0}
\end{align}
\\

\noindent We introduce a finite tunnel coupling between the levels
in both dots described by the Hermitian matrix $\mathbf{T}$, which
for simplicity \cite{Cohen:1977} we assume to be purely
non-diagonal

\begin{equation}
\mathbf{T} = \begin{pmatrix} 0 & t_{12} \\ t_{21} & 0
\end{pmatrix}  \text{, \ }t_{12}=t_{21}^{*}\text{, \
}t_{21}=\left| t_{21}\right| \text{e}^{i\varphi }
\end{equation}
\\

\noindent One obtains a new Hamiltonian, $\mathbf{H} =
\mathbf{H_0} + \mathbf{T}$, with delocalized eigenstates $\left|
\psi_{S}\right\rangle$ (symmetric state) and $\left|
\psi_{A}\right\rangle$ (anti-symmetric state), and eigenvalues
$E_{S}$ and $E_{A}$

\begin{align}
& \mathbf{H} \left| \psi _{S}\right\rangle =  E_{S}\left| \psi
_{S}\right\rangle \nonumber \\
& \mathbf{H} \left| \psi _{A}\right\rangle =  E_{A}\left| \psi
_{A}\right\rangle .
\end{align}

\noindent The new eigenvalues can be expressed in terms of the
eigenvalues of the uncoupled double dot and the tunnel matrix
elements as follows

\begin{align}
& E_S  = E_{M}-\sqrt{\frac{1}{4}(\Delta E)^{2}+|t_{12}|^{2}} \nonumber \\
& E_A  = E_{M}+\sqrt{\frac{1}{4}(\Delta E)^{2}+|t_{12}|^{2}}
\label{eqH}
\end{align}
\\

\noindent where $E_{M} = \frac{1}{2}(E_1 + E_2)$ and $\Delta E =
E_1 - E_2$ and $|t_{12}| = |t_{21}|$. The eigenstates $\left|
\psi_{S}\right\rangle$ and $\left| \psi_{A}\right\rangle$ in the
basis of $\left| \phi _{1}\right\rangle$ and $\left| \phi
_{2}\right\rangle$ are written

\begin{align}
& \left| \psi_{S}\right\rangle  = -  \sin \frac{\theta
}{2}\text{e}^{-i\varphi /2}\left| \phi _{1}\right\rangle +\cos
\frac{\theta }{2}\text{e} ^{i\varphi /2}\left| \phi
_{2}\right\rangle \nonumber \\
& \left| \psi_A\right\rangle  =  \hspace{0.4cm} \cos
\frac{\theta}{2}\text{e}^{-i\varphi /2}\left| \phi
_{1}\right\rangle + \sin \frac{\theta }{2}\text{e} ^{i\varphi
/2}\left| \phi _{2}\right\rangle \label{eigenstates}
\end{align}
\\

\noindent with $\tan \theta = 2|t_{12}| / \Delta E$. Figure
\ref{2levelfig}(b) shows the eigenenergies of the coupled
two-level system as function of $\Delta E$. The renormalized
energy difference, $\Delta E^{\ast}$, is given by

\begin{equation}
\Delta E^{\ast }=E_A-E_S=\sqrt{(\Delta E)^{2}+(2|t_{12}|)^{2}}.
\label{eq11}
\end{equation}

\noindent Note that the effect of the coupling is stronger for
small $\Delta E$, i$.$e$.$ close to the crossing of the
unperturbed energies $E_1$ and $E_2$. Where $E_1$ and $E_2$ cross
($\Delta E$ = 0), we have an anti-crossing of $E_A$ and $E_S$,
with $E_A - E_S$ = $2|t_{12}|$, the minimum bonding--anti-bonding
energy difference. For large $\Delta E$, the eigenenergies of the
coupled double dot approach the eigenenergies of the uncoupled
dots, $E_1$ and $E_2$. The general solution of the {\it
time-dependent} Schr\"{o}dinger equation can be written in the
form

\begin{equation}
\left| \psi (t)\right\rangle =\lambda e^{-iE_At/\hbar}\left| \psi
_{A}\right\rangle +\mu e^{-iE_St/\hbar}\left|
\psi_{S}\right\rangle . \label{eqtimedependent}
\end{equation}
With Eq$.$ (\ref{eigenstates}), $\left| \psi (t)\right\rangle$ can
be expressed in terms of $\left| \phi _{1}\right\rangle $ and
$\left| \phi _{2}\right\rangle $. Since $\left| \phi
_{1}\right\rangle $ and $\left| \phi _{2}\right\rangle $ are not
eigenstates of the total Hamiltonian $\mathbf{H}$, they are no
longer stationary states. If the system is in state $\left| \phi
_{1}\right\rangle $ at time $t$ = 0 ($\left| \psi (0)\right\rangle
=\left| \phi _{1}\right\rangle $) the probability $P_{12}(t)$ of
finding it in the state $\left| \phi _2\right\rangle $ at time $t$
is

\begin{equation}
P_{12}(t)=|\langle \phi _2 | \psi (t) \rangle|^2
\frac{4|t_{12}|^{2}}{4|t_{12}|^{2}+(\Delta E)^{2}}\sin ^{2}\left[
\sqrt{(\Delta E)^{2}+(2|t_{12}|)^{2}}\frac{t}{2\hbar }\right].
\label{eqRabi}
\end{equation}
\\

\noindent Equation (\ref{eqRabi}) describes a coherent charge
oscillation in the double dot system.

\subsection{Photon assisted tunneling in weakly coupled dots}

\label{weakcPAT}

If the inter-dot coupling is weak, electrons are strongly
localized on the individual dots. In section \ref{linquant} we saw
that we expect a resonant current through the double dot system if
$\mu_L \geq \mu_1 = \mu_2 \geq \mu_R$. If we represent a weakly
coupled double dot by a two-level system, we need the discrete
energy levels $E_1$ and $E_2$ to align within the bias window. We
will only consider the discrete, quantum contribution to the
electrochemical potentials and therefore simply use the discrete
level notation $E_1$ and $E_2$ instead of $\mu_1$ and $\mu_2$.\\
\indent An additional time-varying potential $V_{ac}$cos$(2 \pi f
t)$ can induce {\it inelastic} tunnel events when electrons
exchange photons of energy $hf$ with the oscillating field
(frequency, $f$, is typically 1-75 GHz in our experiments). This
inelastic tunneling with discrete energy exchange is known as
photon assisted tunneling (PAT)
\cite{Kouwenhoven:1994a,Kouwenhoven:1994b,Blick:1995}. PAT through
a single quantum dot with well resolved discrete 0D-states is
reviewed in Ref. \cite{Wiel:2002}. PAT is an invaluable
spectroscopic tool for studying the energy spectra of quantum
dots. A theoretical study of PAT in double dots is given in
Refs$.$ \cite{Stoof:1996,Hazelzet:2001}. A voltage drop
$V_{ac}$cos$(2 \pi f t)$ across a tunnel barrier modifies the
tunnel rate through the barrier as \cite{Tien:1963}

\begin{equation}
\widetilde{\Gamma}(E) =\sum\limits_{n=-\infty }^{\infty } {\textup
J}_{n}^{2}(\alpha )\Gamma (E+nhf) \label{eq15}
\end{equation}
\\

\begin{figure}[htbp]
  \begin{center}
    \centerline{\epsfig{file=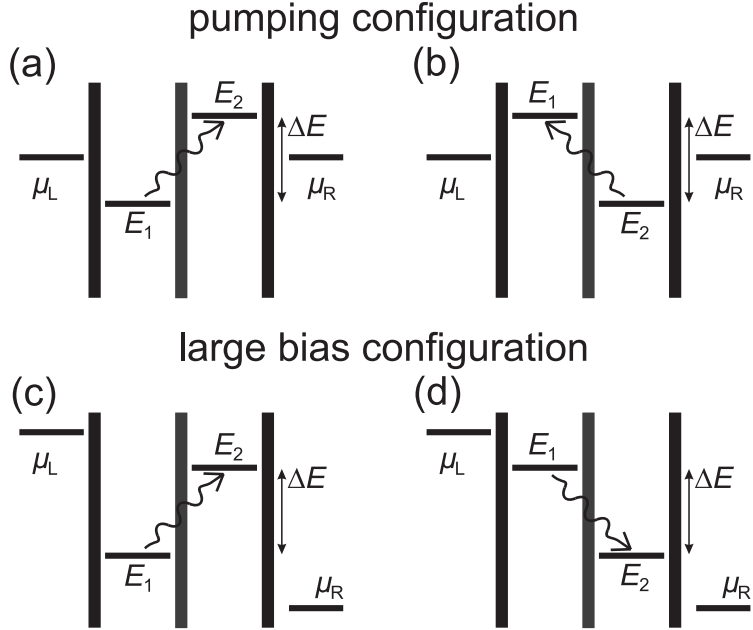, width=10cm, clip=true}}
    \caption{Schematic electrochemical potential diagrams of
    photon assisted tunneling (PAT) in a weakly coupled double
    quantum dot. The upper diagrams (a) and (b) show absorption of
    a photon with energy $hf = \Delta E$ in the so called pumping
    configuration. Although $V=0$, an electron can tunnel from
    left to right through the dot (a), or vice versa (b). The
    lower diagrams show absorption (c) and (stimulated) emission
    (d) of a photon with energy $hf = \Delta E$ in the large bias
    configuration.}
    \label{weakcdiagrams}
  \end{center}
  \vspace{-0.5cm}
\end{figure}

\begin{figure}[htbp]
  \begin{center}
    \centerline{\epsfig{file=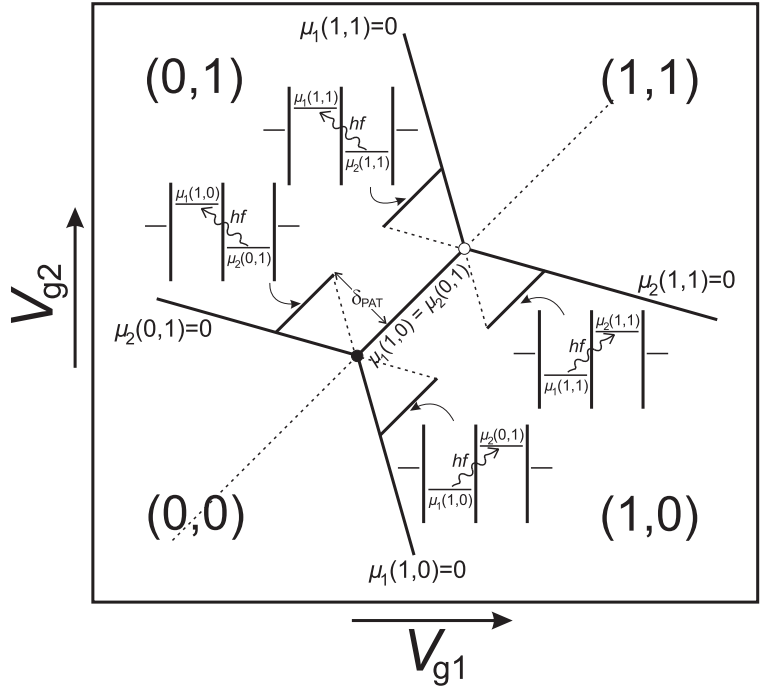, width=10cm, clip=true}}
    \caption{Schematic of a double quantum dot stability diagram in
    the weak coupling and linear transport regime irradiated by
    microwaves with frequency $f$. Next to the triple points, finite
    current is measured along the PAT lines at a distance
    $\delta_{PAT}$ from the line $\mu_1(1,0) = \mu_2(0,1)$, where $\Delta E = hf$.
    The various PAT processes are illustrated by the electrochemical
    potential diagrams.}
    \label{PAThoney}
  \end{center}
  \vspace{-0.5cm}
\end{figure}

\noindent Here $n = 0, \pm 1, \pm 2,...$, and
$\widetilde{\Gamma}(E)$ and $\Gamma (E)$ are the tunnel rates at
energy $E$ with and without an ac voltage, respectively. J$_n
^2(\alpha)$ is the square of the $n$th order Bessel function of
the first kind, evaluated at $\alpha = eV_{ac}/hf$, which
describes the probability that an electron absorbs ($n>0$) or
emits ($n<0$) $n$ photons of
\begin{figure}[htbp]
  \begin{center}
    \centerline{\epsfig{file=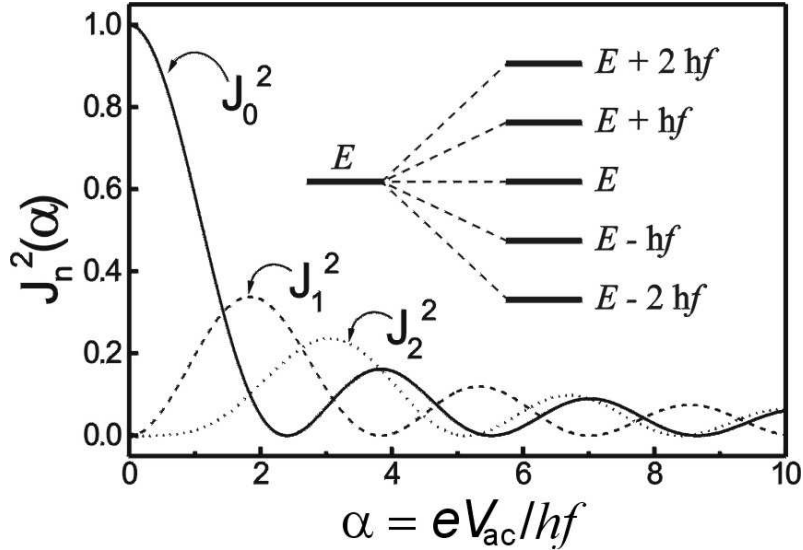, width=10.5cm, clip=true}}
    \caption{Squared Bessel functions of the first kind, J$_n^2(\alpha)$, for $n$ = 0, $\pm$ 1, $\pm$ 2.
    The inset schematically shows the development of sidebands of the original energy as a
    consequence of the microwave field. A positive (negative) $n$ corresponds to the absorption
    (emission) of $n$ photons during the tunnel process. Elastic tunneling corresponds
    to $n$ = 0.}
    \label{besselfig}
  \end{center}
  \vspace{-0.5cm}
\end{figure}
energy $hf$. Thus, the effect of the interaction between a
single-electron state with a classical, oscillating field is that
the energy state $E$ is split in a set of states $E + nhf$ (see
inset to Fig$.$ \ref{besselfig}). The power of PAT as a
spectroscopic tool lies in the fact that PAT can only take place
if the energy difference $\Delta E$ equals an integer number times
the photon energy $hf$: $\Delta E = n hf$, see Fig$.$
\ref{weakcdiagrams}. For the multiple photon processes ($|n| > 1$)
to take place, the microwave power needs to be sufficiently
large.\\
\indent To use PAT as a spectroscopic tool, we can make use of the
configurations shown in Fig$.$ \ref{weakcdiagrams}. In the pumping
configuration \cite{Stafford:1996,Brune:1997}, the double dot is
operated at zero bias voltage. Absorption of a photon with energy
$hf = \Delta E$ leads to pumping of an electron from left to right
(Fig$.$ \ref{weakcdiagrams}(a)) or vice versa (Fig$.$
\ref{weakcdiagrams}(b)). The advantage of this configuration is
that relaxation due to spontaneous emission does not contribute to
the current. Figure \ref{PAThoney} schematically shows how the
honeycomb unit cell of Fig$.$ \ref{honeyelch} changes in the
presence of a microwave field.\\
\indent Alternatively, the double dot can be operated in the large
bias regime as depicted in Fig$.$ \ref{weakcdiagrams}(c),d. In
this regime, in the case of weak coupling with $|t_{12}| \ll
\Delta E, hf, \hbar \Gamma_{L,R}$, the dc PAT current is given by
\cite{Stoof:1996}
\begin{equation}
I_{PAT}=e|t_{12}|^{2}\Gamma _{R}\sum\limits_{n=-\infty }^{\infty }
{\textup J}_{n}^{2}(\alpha )/(\frac{1}{4}\Gamma _{R}^{2}+(n2\pi
f-\Delta E/h)^{2}) \label{eq17}
\end{equation}
\\
The current is composed of a number of satellite peaks, separated
by the photon energy $hf$ and all with width $\Gamma _R$. Note
that the satellite peaks can become of the same order of magnitude
as the main resonance, but that they have a smaller width than the
main resonance. The PAT experiment described below, is performed
on a weakly coupled dot in the large bias regime.
\begin{figure}[htbp]
  \begin{center}
    \centerline{\epsfig{file=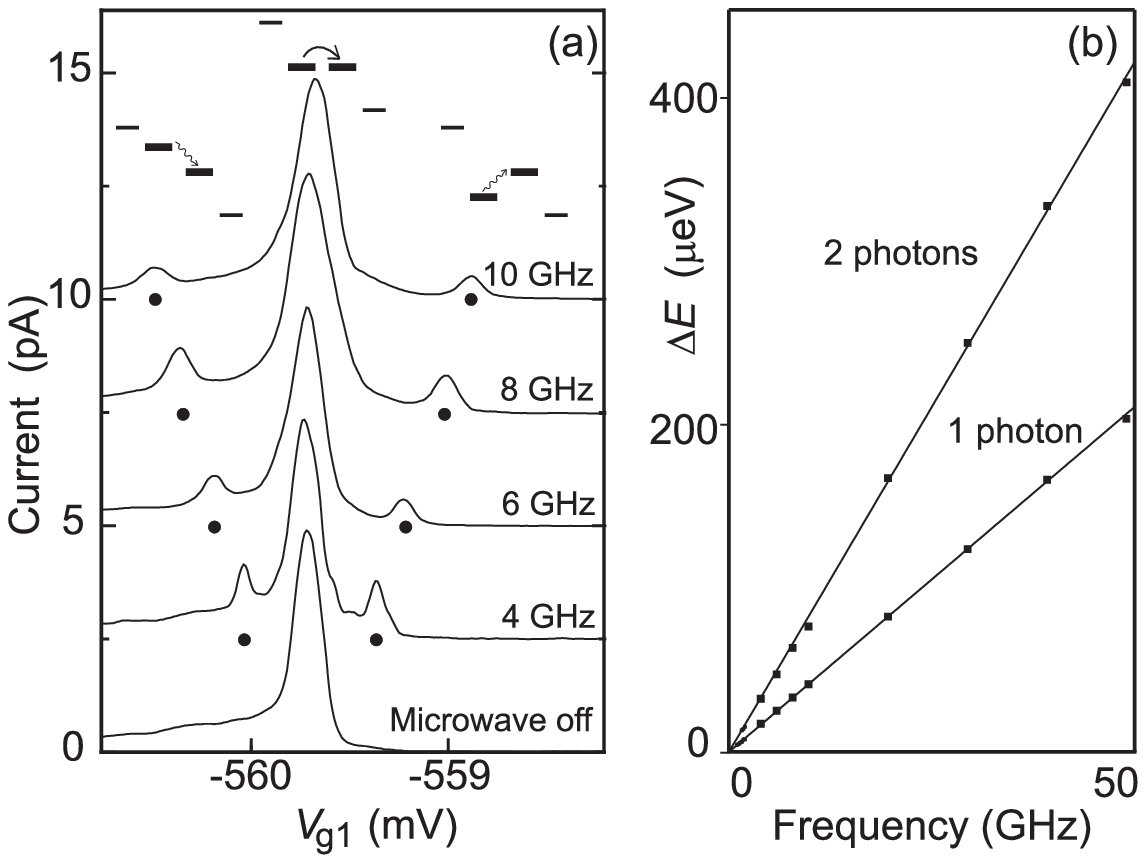, width=14cm, clip=true}}
    \caption{Weakly coupled double quantum dot in the low
    microwave power regime. (a) The upper schematic
    pictures illustrate three configurations of the discrete energy
    level in the left dot relative to the level in the right dot
    (thick solid lines). The electrochemical potentials of the
    leads are indicated by thinner solid lines. The bottom curve
    shows the current as a function of the
    voltage on gate 1, $V_{g1}$, (see Fig$.$ \ref{devices}(b)) for source-drain
    voltage, $V$ = 500 $\mu$V without applying microwaves.
    A single resonance occurs when two levels align. The other
    curves, which have been offset for clarity, show the
    current when microwaves with frequency $f$ from 4 to 10 GHz
    are applied. Now, two additional satellite resonances occur
    when the two levels are exactly a photon energy apart. The
    corresponding photon-assisted tunneling processes are
    illustrated in the upper diagrams.
    (b) Distance between main resonance and first two satellites
    as a function of the applied frequency from 1 to 50 GHz. The
    distance is transferred to energy through $\Delta E = \kappa
    \Delta V_{g1}$ where $\kappa$ is the appropriate capacitance
    ratio for our device that converts gate voltage to energy.
    The agreement between data points and the two solid lines,
    which have slopes of $h$ and 2$h$, demonstrates that we
    observe the expected linear frequency dependence of the one
    and two photon processes.}
    \label{weakcoup1}
  \end{center}
\end{figure}

The double dot, shown in Fig$.$ \ref{devices}(a), is tuned such
that only one level in each dot contributes to electron transport.
The gate voltages are used to shift the level in dot 1 and in dot
2. The resonance in the lowest trace in Fig$.$ \ref{weakcoup1}(a)
arises from the alignment of the two levels. The other traces are
measured while applying a microwave signal. The satellite
resonances are due to PAT processes which involve the emission
(left satellite peak) or absorption (right satellite peak) of one
\begin{figure}[htbp]
  \begin{center}
    \centerline{\epsfig{file=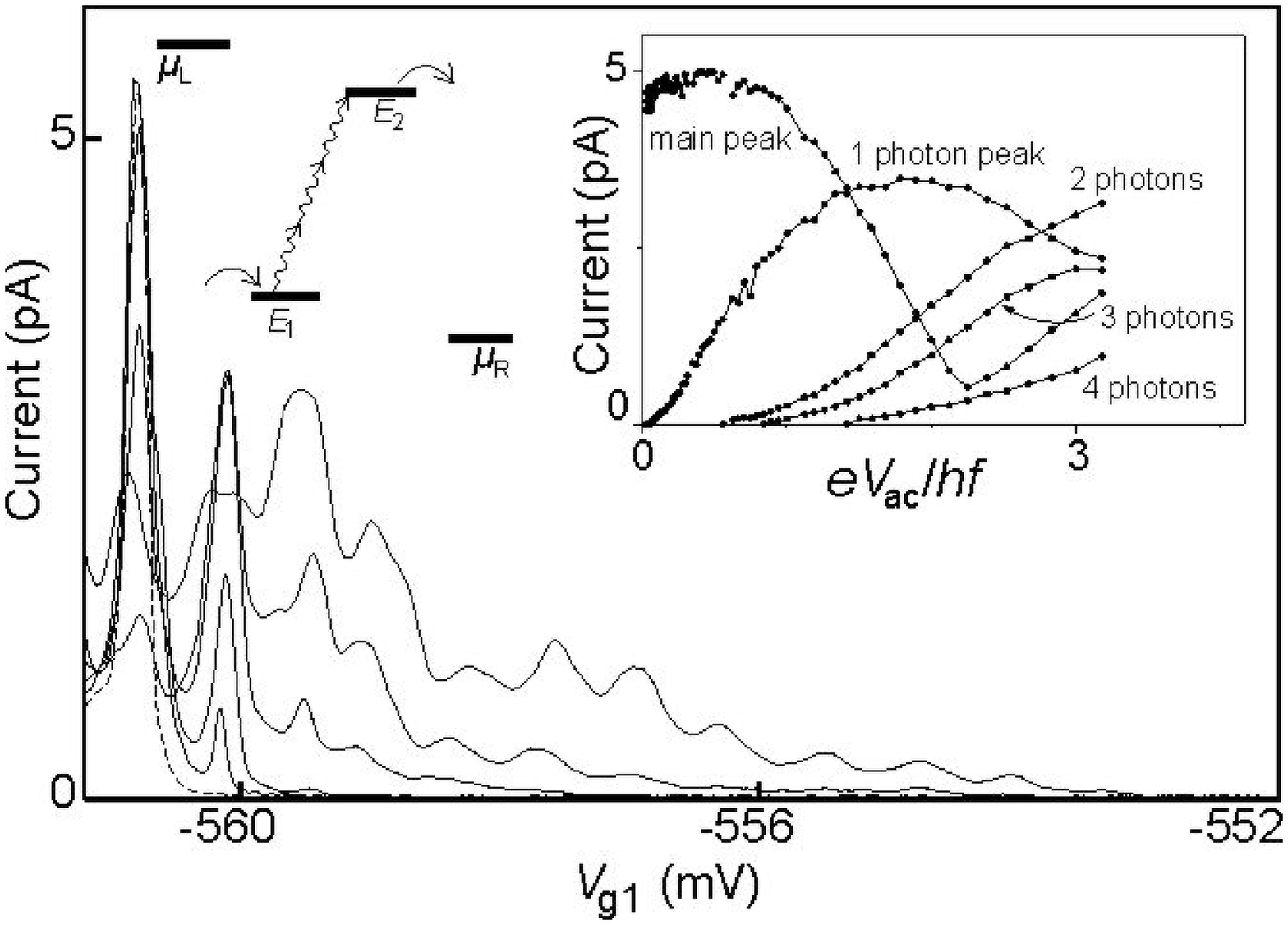, width=10cm, clip=true}}
    \caption{Weakly coupled double quantum dot in the high microwave
    power regime. The main graph shows current versus gate voltage. The dashed curve is without
    microwaves and contains only the main resonance. The solid
    curves are taken at 8 GHz for increasing microwave powers
    resulting in an increasing number of satellite peaks. At the
    right side of the main peak, these correspond to photon
    absorption. $V$ = 700 $\mu$V and the photon energy $hf$ =
    32 $\mu$eV at 8 GHz. At the highest power we observe 11
    satellite peaks, demonstrating multiple photon absorption.
    (left inset) Schematic diagram showing multi-photon
    absorption. (right inset) Height of the first four satellite peaks as a function
    of the microwave amplitude. The observed dependence shows the
    expected Bessel function behavior given in Fig$.$
    \ref{besselfig}.}
    \label{weakcoup2}
  \end{center}
  \vspace{-0.5cm}
\end{figure}
photon. Figure \ref{weakcoup1}(b) shows that the energy separation
of the satellite peaks from the main peak, $\Delta E$, depends
linearly on frequency between 1 and 50 GHz. As we will discuss
below, this linearity implies that the tunnel coupling is
negligible. The electrons are localized on the individual dots and
they have an ionic bonding. The line proportional to $2hf$ is
taken from data at higher microwave powers where electrons absorb
or emit two photons during tunneling.\\
\indent As the microwave power is increased, more satellite peaks
appear corresponding to the absorption of multiple photons, which
are observed up to $n$ = 11 (see Fig$.$ \ref{weakcoup2}). A high
power microwave field strongly perturbs tunneling. This is
reflected by the non-linear dependence of the peak heights on
microwave power. In the right inset to Fig$.$ \ref{weakcoup2} the
peak heights of the main peak and the first four photon satellite
peaks are shown, which agree well with the expected squared Bessel
function behavior shown in Fig$.$ \ref{besselfig}.

\subsection{Photon assisted tunneling in strongly coupled dots}

\label{strongcPAT}

\begin{figure}[htbp]
  \begin{center}
    \centerline{\epsfig{file=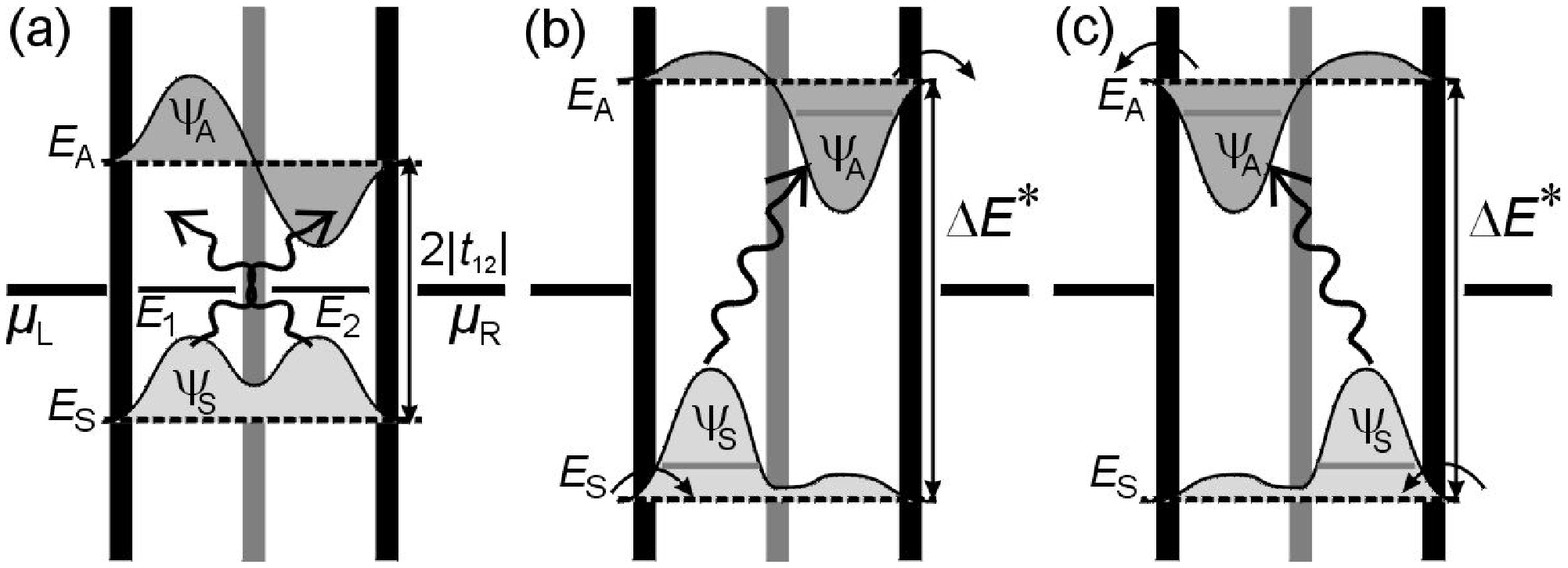, width=12cm, clip=true}}
    \caption{Schematic electrochemical potential diagrams of PAT
    in a strongly coupled double quantum dot in the pumping configuration.
    The diagrams show the symmetric state with wave function $\psi_S$ and energy
    $E_S$ (lower dashed line), and the anti-symmetric state with wave function $\psi_A$ and energy
    $E_A$ (upper dashed line) in combination with the eigenenergies $E_1$ and $E_2$ of the
    weakly coupled double dot (solid lines). (a) $E_1 = E_2 = 0$, $\Delta E^{\ast}= E_A - E_S =
    2|t_{12}|$. Irradiation with photons with energy $hf = 2|t_{12}|$ leads to PAT,
    but the net current through the double dot remains zero. (b) By lowering $-|e|
    \varphi_1$ and increasing $-|e| \varphi_2$ the weight of the wave
    functions is redistributed such that net electron transport from left to right
    occurs. (c) By increasing $-|e| \varphi_1$ and lowering $-|e| \varphi_2$ the weight of the wave
    functions is redistributed such that net electron transport from right to
    left occurs.}
    \label{strongcdiagrams}
  \end{center}
  \vspace{-0.5cm}
\end{figure}

The large bias configuration of Figs$.$ \ref{weakcdiagrams}(c),d
was successfully employed to study PAT in a weakly coupled double
dot system. For the microwave spectroscopy of a strongly coupled
double dot we will make use of the pumping configuration shown in
Fig$.$ \ref{strongcdiagrams}. With increasing the coupling between
the dots, the spontaneous emission rate from the higher level to
the lower one increases as well. The advantage of the pumping
configuration is that these processes are `filtered out' and do
not contribute to the current.

Figure \ref{strongcdiagrams}(a) schematically shows the symmetric
and anti-symmetric states in the double dot for $\Delta E =0$.
When microwave radiation is applied with a frequency such that $hf
= \Delta E^{\ast}= E_A - E_S = 2|t_{12}|$, electrons are pumped
from the left lead to the right lead and vice versa. Since the
weight of the symmetric and anti-symmetric wave function is
distributed equally over both dots (see Fig$.$
\ref{strongcdiagrams}(a)), there is no net current. However, if we
detune the levels ($|\Delta E| > 0$) the weight gets distributed
asymmetrically as shown in Figs$.$ \ref{strongcdiagrams}(b),c and
a net current is generated by applying microwave radiation
matching $hf = \Delta E^{\ast}$. Note that for frequencies $hf <
2|t_{12}|$
no PAT is possible.\\
\indent Multi-photon processes occur if the condition $\Delta E
^{\ast}$ = $nhf$ ($|n| > 1$) is met. Besides allowing for these
higher order photon processes, a high power microwave
\begin{figure}[htbp]
  \begin{center}
    \centerline{\epsfig{file=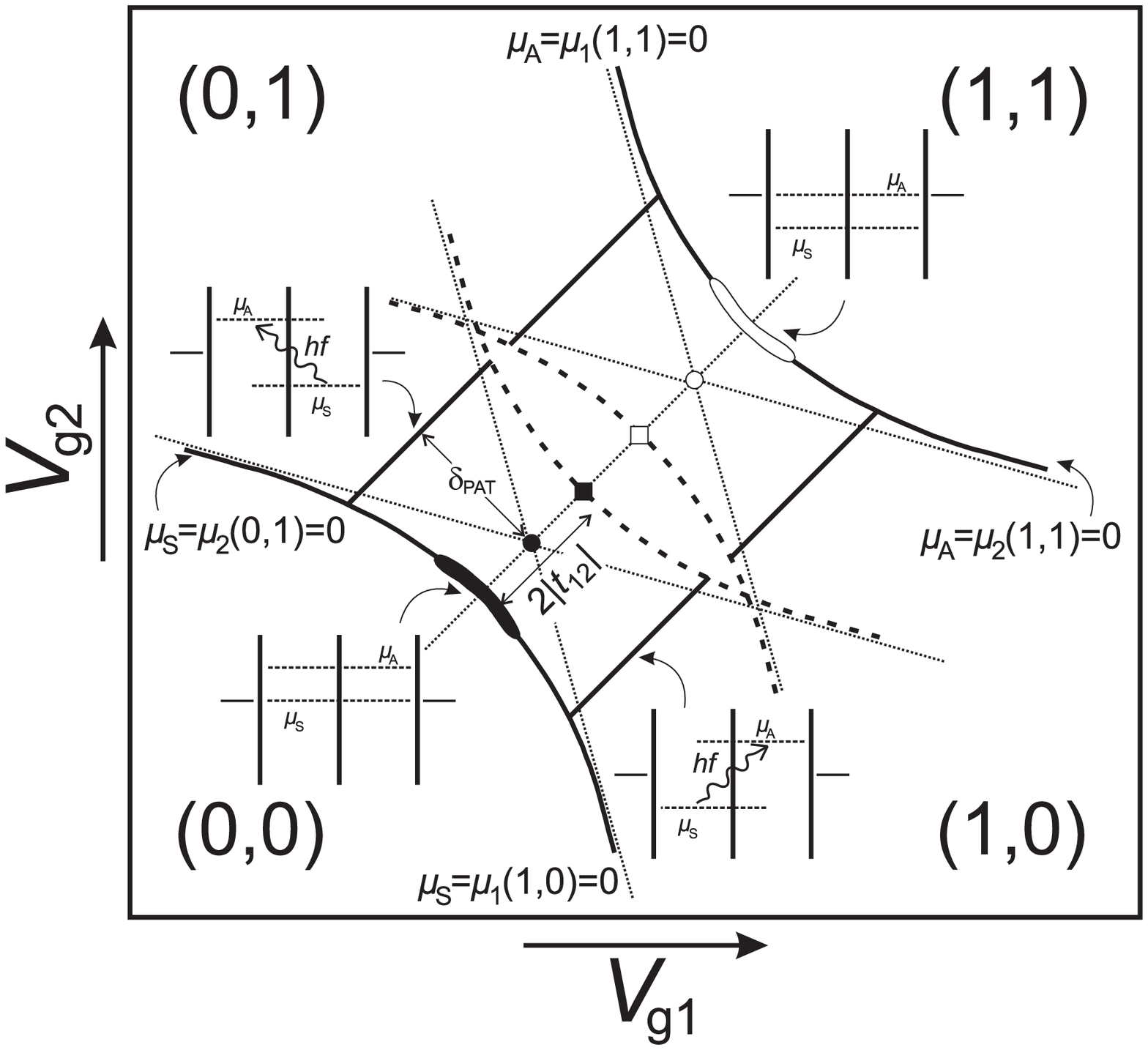, width=10cm, clip=true}}
    \caption{Schematic of a double quantum dot stability diagram in the
    strong coupling and linear transport regime irradiated by microwaves
    with frequency $f$. The symmetric and anti-symmetric states are assumed
    to be singly occupied. At the solid hyperbolic lines either $\mu_S$
    or $\mu_A$ equals zero, marking the separation of charge domains.
    At the dashed hyperbolic lines either $\mu_S$ or $\mu_A$
    equals zero as well, but electron transport is blocked. The triple points of the weakly coupled double dot
    ({\Large $\bullet$} and {\Large $\circ$}) develop in to the black and
    white crescent, respectively. At the position of the black and white square
    no current occurs, as explained in the text. At a distance $\delta_{PAT}$ from the
    dotted line connecting the crescents, $\Delta E^{\ast} = hf$ and PAT occurs. The various configurations
    of the electrochemical potentials are also illustrated.}
    \label{strongchoney}
  \end{center}
  \vspace{-0.5cm}
\end{figure}
field also renormalizes the tunnel coupling to a smaller value.
The energy splitting $\Delta E ^{\ast }$ now becomes
\begin{equation}
\Delta E^{\ast }=\sqrt{(\Delta E)^{2}+[2\mathrm{J}_{0}(\alpha
)|t_{12}|]^{2}}. \label{eqDeltaEstarhp}
\end{equation}

The experiments for strong inter-dot coupling are performed on the
device shown in Fig$.$ \ref{devices}(b). To single out the current
only due to microwaves, we operate the device as an electron pump
driven by photons \cite{Stafford:1996,Brune:1997} (see the
diagrams in Fig$.$ \ref{strongcdiagrams}). An electron is excited
from the bonding to the anti-bonding state if the condition $hf =
\Delta E^{\ast}$ is fulfilled, or conversely
\begin{equation}
\Delta E =\sqrt{(hf)^{2}-(2\mathrm{J}_{0}(\alpha )|t_{12}|)^{2}}.
\label{eqDeltaEhp}
\end{equation}

\begin{figure}[htbp]
  \begin{center}
    \centerline{\epsfig{file=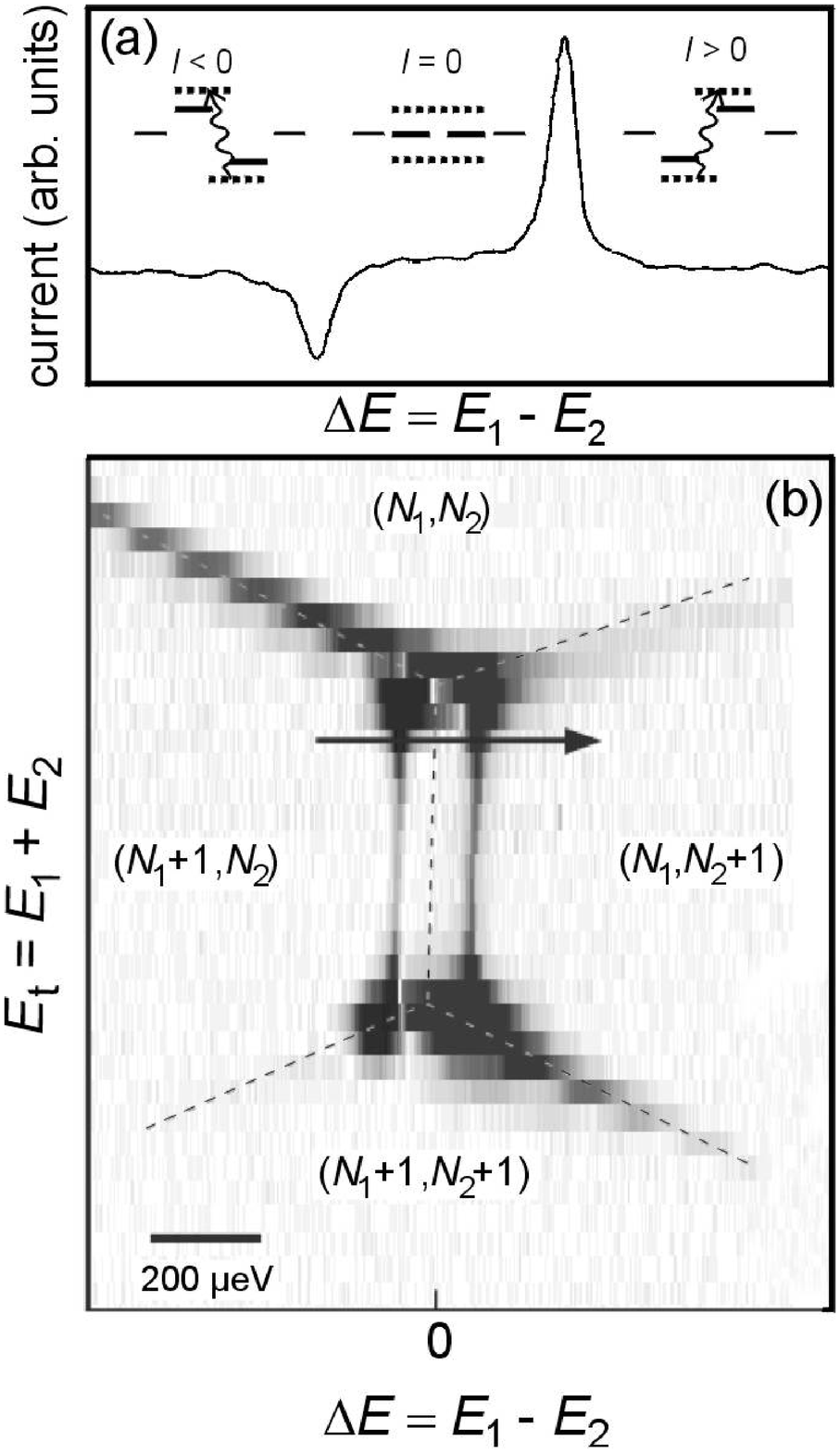, width=8cm, clip=true}}
    \caption{Strongly coupled double dot in the low-power regime.
    (a) Current through the double dot as function of the energy difference
    between the level in the left and the right dot. The current trace
    is taken from the stability diagram in (b) at the position
    indicated by the arrow. The diagrams depict the discrete levels
    $E_1$ and $E_2$ in the two dots for the case that the coupling is
    weak (solid lines) and the bonding and anti-bonding states in
    the case of strong coupling (dotted lines). The PAT processes
    leading to a negative (left diagram) and a positive current (right
    diagram) are indicated. (b) Gray-scale plot of the current
    through the double dot versus the energy level difference, $\Delta E$, and
    the total energy, $E_t$. The bias voltage is 6 $\mu$V and the applied
    microwave frequency is 16 GHz such that $hf$ = 66 $\mu$eV. The dashed lines divide the stability diagram
    in 4 regions of stable electron numbers. In between the two triple points clear features of
    photon-assisted tunneling are seen. The black arrow indicates the
    position of the trace shown in (a).}
    \label{toshipat}
  \end{center}
  \vspace{-0.5cm}
\end{figure}

Figure \ref{strongchoney} schematically shows the stability
diagram for a strongly coupled double dot in the presence of a
microwave field. Here we assume that the symmetric and
anti-symmetric states can only be occupied by a single electron.
In other words, we assume spinless electrons. The triple points of
the weakly coupled double dot, denoted by {\Large $\bullet$} and
{\Large $\circ$}, develop into a black and a white
crescent, respectively. The length of these crescents increases with $|t_{12}|$.\\
\indent Moving along the dotted line connecting the crescents from
lower left to upper right, first the symmetric state aligns with
the electrochemical potentials of the leads (at the black
crescent). Current through the double dot is possible via the
electron transfer process of Fig$.$ \ref{honeycombs}(d). At the
black square, the anti-symmetric state aligns with the leads.
However, current is blocked, since an extra electron is already
added to the double dot and the charging energy $E_{Cm}$ is not
available yet. At the white square, the electrochemical potential
for adding the second electron to the symmetric state aligns with
the leads. As we assumed single occupation of the delocalized
states, current is blocked here as well. When arriving at the
white crescent, the electrochemical potential for adding the
second electron to the double dot in the (empty) anti-symmetric
state becomes available. This enables the hole transfer process of
Fig$.$ \ref{honeycombs}(d).\\
\indent The black bars at a distance $\delta_{PAT}$ from the
dotted line connecting the crescents, denote the places where
$\Delta E^{\ast} = hf$ and PAT occurs. Note that PAT is only
possible if the photon energy exceeds the coupling energy, $hf
\geq 2|t_{12}|$. Depending on the level configuration, pumping
results in a negative or positive contribution to the current, as
shown in Fig$.$ \ref{toshipat}(a) (here we choose $I>0$ for an
electron moving from left to right). Figure \ref{toshipat}(b)
shows a part of the corresponding stability diagram between two
triple points, clearly showing the energy regions of constant
charge and extra transport lines due to PAT.

\begin{figure}[htbp]
  \begin{center}
    \centerline{\epsfig{file=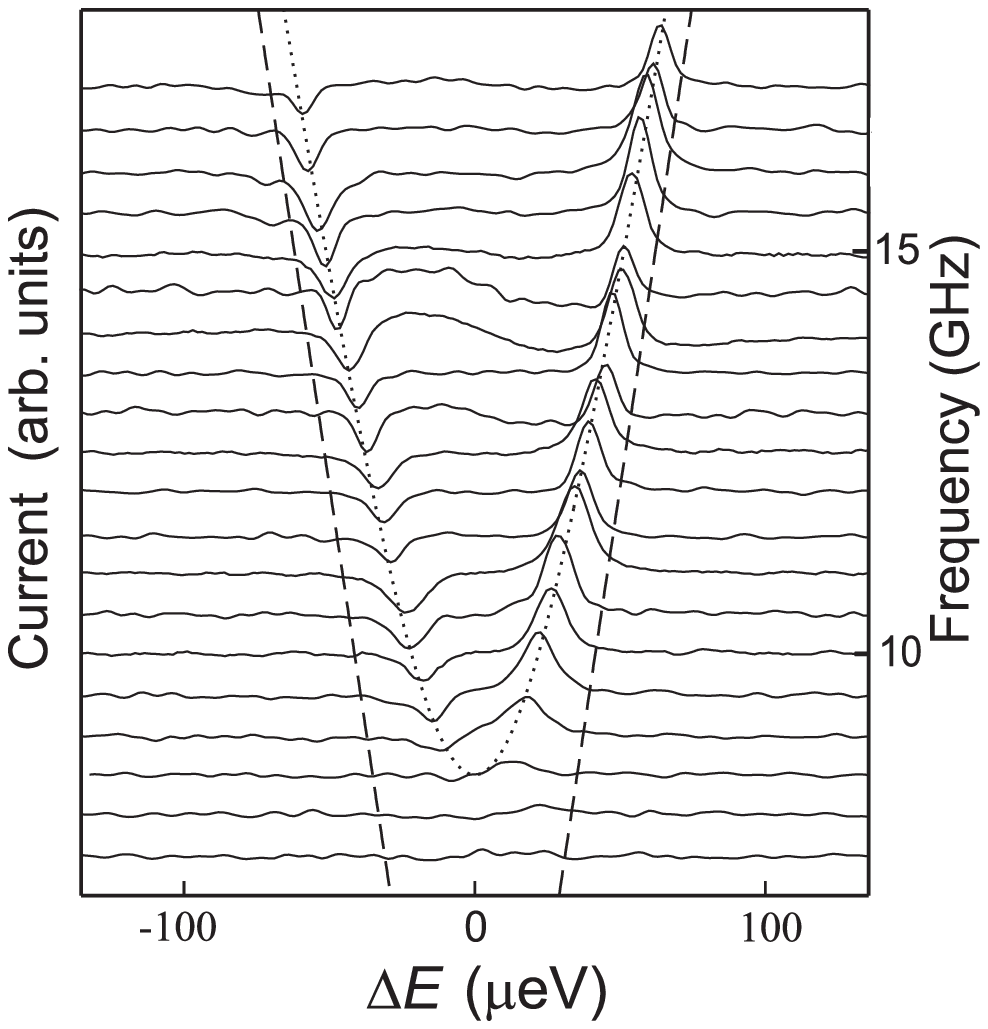, width=9cm, clip=true}}
    \caption{Measured pumped current through the strongly coupled double dot.
    Gates 1 and 3 are swept simultaneously in such a way that we vary the energy
    difference $\Delta E$. The different traces are taken at different
    microwave frequencies, and are offset such that the right vertical
    axis gives the frequency. The main resonance is absent as we have set
    $V$ = 0. The satellite peaks typically have an amplitude of 0.5 pA.
    For weakly coupled dots the satellite peaks are expected to move
    linearly with frequency, thereby following the straight dashed lines.
    In contrast, we observe that the satellite peaks follow the fitted
    dotted hyperbola $hf = [\Delta E^2 + (2|t_{12}|)^2]^{1/2}$ using the coupling
    $|t_{12}|$ as a fitting parameter.}
    \label{patpeaks}
  \end{center}
  \vspace{-0.5cm}
\end{figure}

\begin{figure}[htbp]
  \begin{center}
    \centerline{\epsfig{file=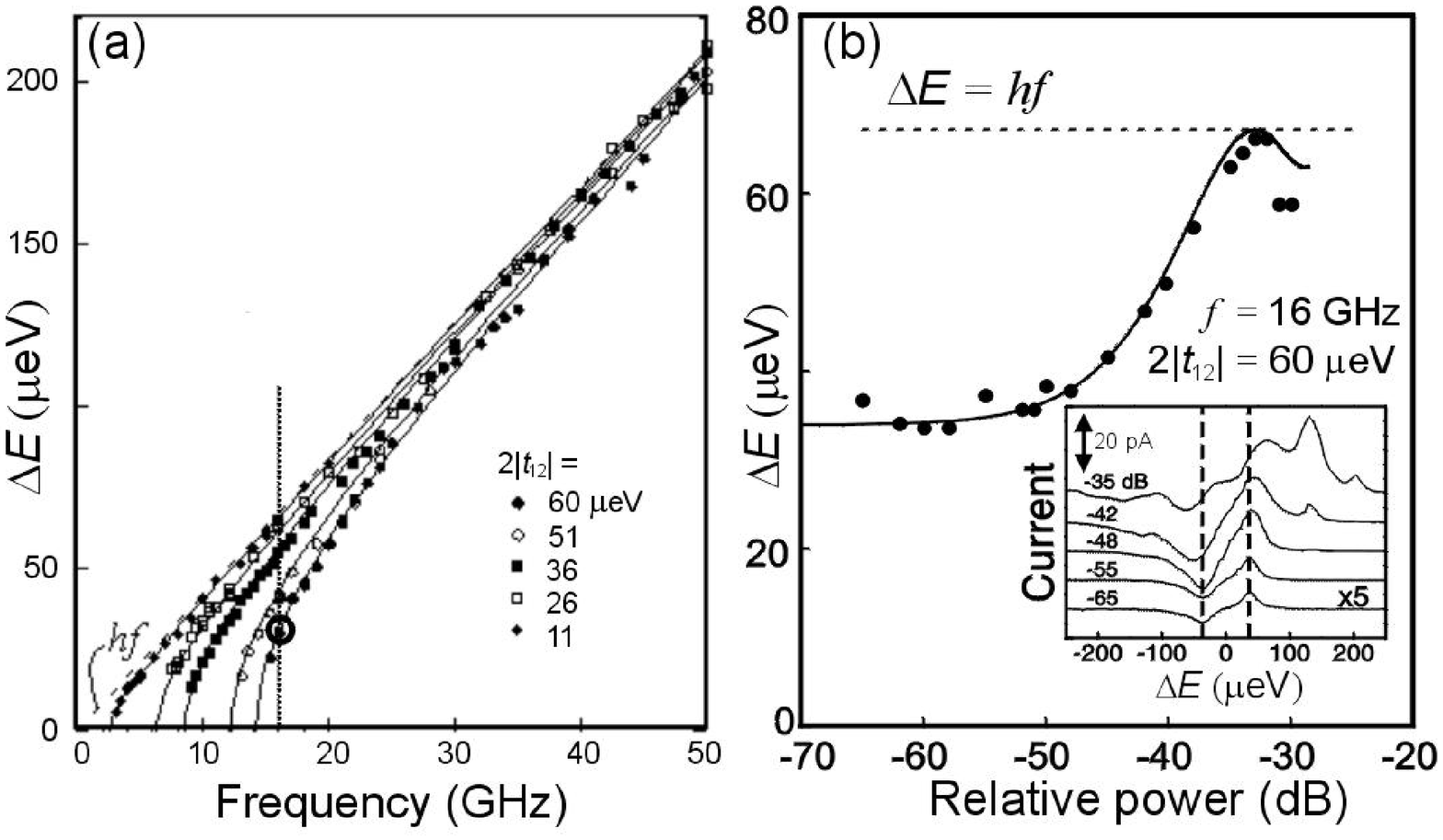, width=14.5cm, clip=true}}
    \caption{(a) Half the spacing in gate voltage between the positive and
    negative satellite peaks as a function of frequency for $e V_{ac} \ll hf$.
    Gate voltage spacing has been
    transferred to energy difference $\Delta E$ (see also figure caption
    Fig$.$ \ref{weakcoup1}(b)). Different curves correspond to different
    tunnel couplings $|t_{12}|$. Solid lines are theoretical fits to
    Eq$.$ (\ref{eqDeltaElp}). In the limit of weak coupling,
    this reduces to $\Delta E = hf$, which is indicated by the dashed line.
    The resulting values for 2$|t_{12}|$ are given in the figure. The coupling is
    varied by applying different voltages to the center gate or by
    changing the magnetic field ($\blacklozenge$ $B$ = 3.3 T; $\blacksquare$
    $B$ = 2.2 T; other curves $B$ = 0 T). The circle marks a coupling of 60
    $\mu$eV and frequency of 16 GHz (dotted line). (b) Strongly coupled
    double dot (2$|t_{12}|$ = 60 $\mu$eV) in the high microwave power regime for
    $f$ = 16 GHz (along dotted line in (a)). The inset shows the measured PAT current as a function of
    $\Delta E$ for different powers. The positions of the PAT peaks at the lowest power
    are indicated with two dashed lines. The PAT peak separation becomes larger
    for increasing microwave power. For higher
    powers, multi-photon processes can also take place, which result in extra
    current peaks. In the main part,
    half the PAT peak separation in energy as function of the relative microwave
    power is shown. The solid line is a fit to Eq$.$ (\ref{eqDeltaEhp}), $f$ = 16 GHz,
    2$|t_{12}|$ = 60 $\mu$eV. Because of the relative power scale, the fitting curve
    has been adjusted horizontally to obtain the best fit.}
    \label{strongcouphp}
  \end{center}
  \vspace{-0.5cm}
\end{figure}

Figure \ref{patpeaks} shows measured current traces as a function
of the uncoupled energy splitting $\Delta E$, where from top to
bottom the applied microwave frequency is decreased from 17 to 7.5
GHz in 0.5 GHz steps. At the highest frequencies, the distance
between the pumping peaks is close to 2$\Delta E$. However, the
peak distance decreases faster than linearly as the frequency is
lowered; in fact the peaks follow the dotted hyperbola rather than
the dashed straight lines. The distance goes to zero when the
frequency approaches the minimum energy gap between bonding and
anti-bonding states, $hf = 2|t_{12}|$.
\noindent The coupling between the dots can be decreased by
changing the gate voltage on the center gate to more negative
values, or by applying a magnetic field perpendicular to the
sample. In Fig$.$ \ref{strongcouphp}(a) we plot half the spacing
between the positive and negative satellite peaks as a function of
frequency. The microwave power is kept as low as possible in order
to meet the condition $eV_{ac} \ll hf$. In that case
J$_0^2(\alpha) \approx 1$ and the general relation Eq$.$
(\ref{eqDeltaEhp}) reduces to

\begin{equation}
\Delta E =\sqrt{(hf)^{2}-(2|t_{12}|)^{2}} \label{eqDeltaElp}.
\end{equation}
\\
Different symbols correspond to different center gate voltage
settings and magnetic fields. The solid lines are fits to Eq$.$
(\ref{eqDeltaElp}). The good agreement with Eq$.$
(\ref{eqDeltaElp}) demonstrates the control over the formation of
a covalent bonding between the two dots and that the condition
$eV_{ac} \ll hf$ is satisfied.

We now discuss the case $eV_{ac} \gtrsim hf$. As can be seen in
Fig$.$ \ref{besselfig}, J$_0^2(\alpha)$ deviates from 1 in this
case and cannot be neglected as before \cite{Wiel:1999}. In Fig$.$
\ref{strongcouphp}(b) we show the power dependence for the case of
a coupling of 60 $\mu$eV and a microwave frequency of 16 GHz, as
indicated by the circle in Fig$.$ \ref{strongcouphp}(a) (similar
results have been obtained for other couplings and microwave
frequencies). The inset to Fig$.$ \ref{strongcouphp}(b) shows the
measured PAT current as a function of $\Delta E$ for different
powers. The absolute value of the microwave power at the position
of the double quantum dot is unknown. Therefore, we use a relative
microwave power scale, which is expressed in terms of the
attenuation of the microwave source signal. The positions of the
PAT peaks at the lowest power are indicated with two dashed lines.
Increasing the microwave power from the lowest value, the PAT peak
separation becomes larger, which is in agreement with Eq$.$
(\ref{eqDeltaEhp}). For higher powers, multi-photon processes can
also take place, which result in extra current peaks. Figure
\ref{strongcouphp}(b) shows half the PAT peak separation energy as
function of the relative microwave power. The solid line is a fit
with Eq$.$ (\ref{eqDeltaEhp}), $f$ = 16 GHz, $2|t_{12}|$ = 60
$\mu$eV. Because of the relative power scale, the fitting curve
has been adjusted horizontally to give the best fit. We thus see
that the microwave power effectively reduces the coupling between
the dots. This is further illustrated by the vertical dotted line
in Fig$.$ \ref{strongcouphp}(a) at $f$ = 16 GHz. At -33 dB the
energy separation equals $hf$, which implies that the higher
microwave power has effectively reduced the coupling between the
dots. We can obtain an estimate of the power by noting that
J$_0^2(\alpha)$ has its first zero for $\alpha = eV_{ac}/hf$ = 2.4
and hence $V_{ac}$ = 0.16 mV.

\section{CONCLUSIONS}

By coupling two quantum dots in series, we obtain a system with
fundamentally different behavior and possibilities in comparison
to a single quantum dot. In this review we have discussed the
superiority of a double quantum dot system in determining the
intrinsic lifetime of quantum states and in probing intra-dot
level repulsion. Next to the added value as a spectroscopic
instrument, the double dot manifests itself as an artificial
molecule. By changing the inter-dot coupling, we have been able to
tune the double dot from an ionic-like bonded to a covalent-like
bonded molecule.\\
\indent Now that the ability to create and manipulate double
quantum dots has been shown, the next challenge lies in the study
and time-control of coherent phenomena in these systems. Double
quantum dots have been suggested as possible candidates for
building blocks of a quantum computer \cite{Loss:1998}. We have
shown that it is indeed possible to coherently couple dots, and
that one can induce transitions between the extended states. The
next crucial step towards quantum logic gates is to show that the
coherence of the superposition is preserved on time scales much
longer than the time needed for manipulating the electron wave
functions. The time-resolved measurement of coherent charge
oscillations in double quantum dots using pulsed gate voltages,
will be an essential step in determining the dephasing time in
these systems.\\
\indent In addition, the role of the electron spin and its
relaxation and dephasing times ($T_1$ and $T_2$, respectively)
need to be characterized, as they are essential in the proposed
quantum bit schemes based on coupled electron spins in double dots
\cite{Loss:1998}. The relaxation time $T_1$ in quantum dots has
been experimentally shown to exceed 10 - 100 $\mu$s if the
relaxation involves a spin-flip \cite{Fujisawa:2002}. The
determination of the spin dephasing time $T_2$, which is the most
relevant timescale for quantum computing purposes, will be an
experimental challenge for the near future. We conclude that our
work on double dots so far, in combination with the results on the
spin relaxation times in this kind of systems, forms a promising
point of departure for further study on the suitability of double
quantum dots as quantum coherent devices.

\section*{Acknowledgments}

We would like to thank R. Aguado, S.M. Cronenwett, D.C. Dixon, S.
Godijn, P. Hadley, C.J.P.M. Harmans, R.V. Hijman, Y. Hirayama, K.
Ishibashi, M.P. Janus, K.K. Likharev, F. Mallens, J.E. Mooij,
Yu.V. Nazarov, T.H. Oosterkamp, R.M. Schouten, T.H. Stoof, M.J.
Uilenreef, N.C. van der Vaart and L. Vandersypen for their help.
We acknowledge financial support from the DARPA grant number
DAAD19-01-1-0659 of the QuIST program, the Specially Promoted
Research Grant-in-Aid for Scientific Research; the Ministry of
Education, Culture, Sports, Science and Technology in Japan; the
Dutch Organization for Fundamental Research on Matter (FOM); the
Core Research for Evolutional Science and Technology (CREST-JST);
and the European Union through a Training and Mobility of
Researchers (TMR) Program network.

\begin{appendix}

\section*{Electrostatic energy of quantum dots}

In this appendix we derive the electrostatic energy of a single
and double quantum dot system. Before addressing these specific
systems, we briefly discuss the method followed \footnote{For a
discussion of the electrostatics of a charging network see
http://qt.tn.tudelft.nl/$\sim$ hadley/set/electrostatics.html.}.

\subsection{Electrostatics of a system of $N$ conductors}

Consider a system consisting of $N$ conductors. A capacitance can
be defined between each conductor and every other conductor as
well as a capacitance from each of the $N$ conductors to ground.
This results in a total of $N(N + 1)/2$ capacitors. The capacitor
between node $j$ and node $k$ has a capacitance $c_{jk}$ and
stores a charge $q_{jk}$. The total charge on node $j$ is the sum
of the charges on all of the capacitors connected to node $j$
\begin{equation}
Q_j = \sum_{k=0}^N q_{jk} = \sum_{k=0}^N c_{jk}(V_j - V_k)
\end{equation}
Here $V_j$ is the electrostatic potential of node $j$ and ground
is defined to be at zero potential, $V_0 = 0$. The charges on the
nodes are linear functions of the potentials of the nodes so this
can be expressed more compactly in matrix form
\begin{equation}
\overrightarrow{Q} = \textbf{C} \overrightarrow{V}
\end{equation}
where \textbf{C} is called the capacitance matrix. A diagonal
element of the capacitance matrix, $C_{jj}$, is the total
capacitance of node $j$
\begin{equation}
C_{jj} = \sum_{k=0,k \neq j}^N c_{jk}
\end{equation}
An off-diagonal element of the capacitance matrix is minus the
capacitance between node $j$ and node $k$, $C_{jk} = C_{kj} = -
c_{jk}$. The electrostatic energy of this system of conductors is
the sum of the electrostatic energy stored on the $N(N+1)/2$
capacitors and can be conveniently expressed using the capacitance
matrix
\begin{equation}
U = \frac{1}{2} \overrightarrow{V} \cdot \textbf{C}
\overrightarrow{V} = \frac{1}{2} \overrightarrow{V} \cdot
\overrightarrow{Q} = \frac{1}{2} \overrightarrow{Q} \cdot
\textbf{C}^{-1} \overrightarrow{Q} \label{Appenergy}
\end{equation}

Voltage sources can be included in the network by treating them as
nodes with large capacitances to ground and large charges on them
such that $V = Q/C$. In this case, it is numerically difficult to
compute the inverse of the capacitance matrix since it contains
large elements. However, it is not necessary to invert the entire
capacitance matrix since the voltages on the voltage sources are
already known. Only the voltages on the other nodes need to be
determined. These voltages can be determined by writing the
relation between the charges and the voltages as
\begin{equation}
\begin{pmatrix} \overrightarrow{Q}_c \\ \overrightarrow{Q}_v
\end{pmatrix} = \begin{pmatrix} \mathbf{C_{cc}} & \mathbf{C_{cv}}
\\ \mathbf{C_{vc}} & \mathbf{C_{vv}} \end{pmatrix} \begin{pmatrix} \overrightarrow{V}_c \\ \overrightarrow{V}_v
\end{pmatrix}
\label{submatrices}
\end{equation}
Here $\overrightarrow{Q}_c$ and $\overrightarrow{V}_c$ are the
charges and the voltages on the charge nodes,
$\overrightarrow{Q}_v$ and $\overrightarrow{V}_v$ are the charges
and the voltages on the voltage sources, and the capacitance
matrix has been expressed in terms of four sub-matrices. The
voltages on the charge nodes are then
\begin{equation}
\overrightarrow{V}_c =
\mathbf{C_{cc}^{\normalfont{-1}}}(\overrightarrow{Q}_c -
\mathbf{C_{cv}} \overrightarrow{V}_v) \label{voltage}
\end{equation}
and the electrostatic energy can be calculated with Eq$.$
(\ref{Appenergy}).

\subsection{Single quantum dot}

\begin{figure}[htbp]
  \begin{center}
    \centerline{\epsfig{file=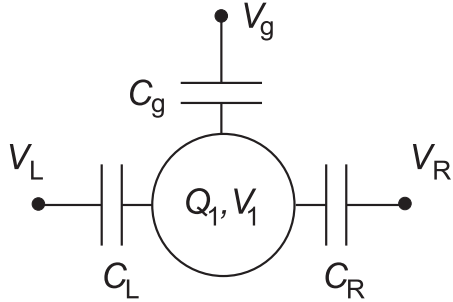, height=4cm, clip=true}}
    \caption{Network of capacitors and voltage nodes used to calculate the
    electrostatic energy of a single quantum dot.}
    \label{AppSD}
  \end{center}
\end{figure}

\noindent We write the total charge $Q_1$ on the dot as the sum of
the charges on all the capacitors connected to the dot (see Fig$.$
\ref{AppSD})
\begin{align}
& Q_1 = C_L (V_1 - V_L) + C_g (V_1 - V_g) + C_R (V_1 - V_R)
\Rightarrow \nonumber \\
& Q_1 + C_L V_L + C_g V_g + C_R V_R = C_1 V_1
\end{align}
where $C_1$ is the total capacitance coupled to the dot, $C_1 =
C_L + C_g + C_R$. The capacitance matrix $\mathbf{C_{cc}}$ only
has one element. Using Eq$.$ (\ref{Appenergy}) and substituting
$Q_1 = - (N_1 - N_0) |e|$, we find
\begin{equation}
U(N_1) = \frac{[-(N_1 - N_0) |e| + C_L V_L + C_g V_g + C_R V_R
]^2}{2 C_1}
\end{equation}
where $N_0$ is the number of electrons on the dot when all voltage
sources are zero, which compensates the positive background charge
originating from donors in the heterostructure.

\subsection{Double quantum dot}

\begin{figure}[htbp]
  \begin{center}
    \centerline{\epsfig{file=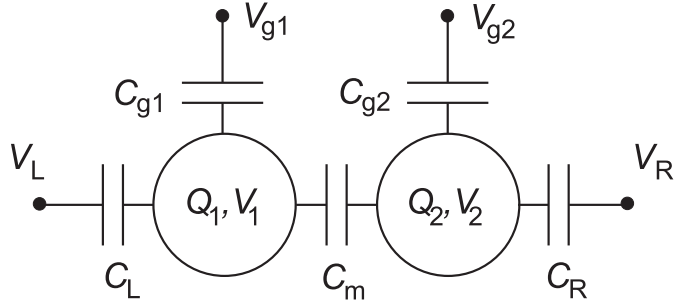, height=4cm, clip=true}}
    \caption{Network of capacitors and voltage nodes used to calculate the
    electrostatic energy of a double quantum dot.}
    \label{AppDD}
  \end{center}
\end{figure}

\noindent We write the total charge $Q_{1(2)}$ on dot 1(2) as the
sum of the charges on all the capacitors connected to dot 1(2)
(see Fig$.$ \ref{AppDD})
\begin{align}
& Q_1 = C_L (V_1 - V_L) + C_{g1} (V_1 - V_{g1}) + C_m (V_1 - V_2) \nonumber \\
& Q_2 = C_R (V_2 - V_R) + C_{g2} (V_2 - V_{g2}) + C_m (V_2 - V_1)
\end{align}
We can write this as
\begin{equation}
\begin{pmatrix} Q_1 + C_L V_L + C_{g1} V_{g1} \\ Q_2 + C_R V_R + C_{g2} V_{g2}
\end{pmatrix} = \begin{pmatrix} C_1 & -C_m
\\ -C_m & C_2 \end{pmatrix} \begin{pmatrix} V_1 \\ V_2
\end{pmatrix}
\end{equation}
where $C_2 = C_R + C_{g2} + C_m$. The above expression in the form
of Eq$.$ (\ref{voltage}) reads
\begin{equation}
\begin{pmatrix} V_{1} \\ V_{2} \end{pmatrix} = \frac{1}{C_1 C_2 - C_m^2} \begin{pmatrix} C_2 & C_m
\\ C_m & C_1 \end{pmatrix} \begin{pmatrix} Q_1 + C_L V_L + C_{g1} V_{g1} \\ Q_2 + C_R V_R + C_{g2} V_{g2}
\end{pmatrix}
\end{equation}
The electrostatic energy of the double dot system can now be
calculated using Eq$.$ (\ref{Appenergy}). For the case $V_L = V_R
= 0$ and $Q_{1(2)} = - N_{1(2)} |e|$ this becomes
\begin{align}
U(N_1,N_2) = & \hspace{0.2cm} \frac{1}{2} N_1^2 E_{C1} +
\frac{1}{2} N_2^2 E_{C2}
+ N_1 N_2 E_{Cm} + f(V_{g1},V_{g2}) \label{AppDDenergy} \\
f(V_{g1},V_{g2}) = & \hspace{0.2cm} \frac{1}{-|e|}\{C_{g1}
V_{g1}(N_1 E_{C1} + N_2 E_{Cm}) +  C_{g2} V_{g2}(N_1 E_{Cm} + N_2
E_{C2}) \} \nonumber
\\
& \hspace{0.2cm} + \frac{1}{e^2}\{\frac{1}{2} C^2_{g1}
V^2_{g1}E_{C1} + \frac{1}{2} C^2_{g2} V^2_{g2} E_{C2} + C_{g1}
V_{g1} C_{g2} V_{g2} E_{Cm} \} \nonumber
\end{align}
with
\begin{equation}
E_{C1}=e^{2}\frac{C_{2}}{C_{1}C_{2}-C_{m}^{2}}; \
E_{C2}=e^{2}\frac{C_{1}}{C_{1}C_{2}-C_{m}^{2}}; \
E_{Cm}=e^{2}\frac{C_m}{C_{1}C_{2}-C_{m}^{2}} \label{AppEC}
\end{equation}
\\

\end{appendix}

\bibliographystyle{apsrmp}


\newpage


\end{document}